\documentclass[notitlepage,aps,12pt,tightenlines,nofootinbib,showkeys]{revtex4}
\pdfoutput=1

\usepackage{amsmath}
\usepackage{amssymb,amsfonts}
\usepackage{bm}
\usepackage[mathcal]{euscript}
\usepackage{graphicx}
\usepackage{subfigure}
\usepackage{color}
\usepackage{verbatim}
\usepackage{hyperref}

\graphicspath{{figs/}{figs_lo/}}

\newcommand{\commentjlt}[1]{}

\newcommand{\mathnotation}[2]{\newcommand{#1}{\ensuremath{#2}}}

\newcommand{\ie}{\textit{i.e.}}
\newcommand{\etal}{\textit{et~al.}}

\renewcommand{\l}{\left}			% \left
\renewcommand{\r}{\right}			% \right
\newcommand{\pos}[1]{#1^+}
\renewcommand{\neg}[1]{#1^-}
\mathnotation{\ldef}{\mathrel{\raisebox{.069ex}{:}\!\!=}}% Left define
\mathnotation{\rdef}{\mathrel{=\!\!\raisebox{.069ex}{:}}}% Right define
\mathnotation{\tim}{t}				% Time
\mathnotation{\nn}{n}				% # of strands
\mathnotation{\Br}{B}				% Braid group
\mathnotation{\R}{R}				% Domain
\mathnotation{\ip}{i}				% Counter for braid elements
\mathnotation{\jp}{j}				% Counter for braid elements
\mathnotation{\per}{m}				% iterates of periodic orbit
\mathnotation{\htop}{h}				% Topological entropy
\newcommand{\htopb}[1]{\htop_{\mathrm{braid}}^{(#1)}}
\mathnotation{\htopf}{\htop_{\mathrm{flow}}}	% Topological entropy of flow
\mathnotation{\ac}{a}				% Dynnikov coord a
\mathnotation{\bc}{b}				% Dynnikov coord b
\mathnotation{\cc}{c}				% Dynnikov coord c
\mathnotation{\dc}{d}				% Dynnikov coord d
\mathnotation{\fc}{f}				% Dynnikov coord f
\mathnotation{\acnew}{\ac'}
\mathnotation{\bcnew}{\bc'}
\mathnotation{\abv}{\bm{u}}			% Dynnikov coord vector
\mathnotation{\Nint}{L}				% Intersection #
\mathnotation{\Nreal}{N_{\mathrm{real}}}	% Number of realizations
\mathnotation{\dt}{\Delta\tim}			% Length of time intervals
\mathnotation{\ti}{q}				% Label of time intervals
\mathnotation{\tmax}{\tim_{\mathrm{max}}}	% Max integration time

\newtheorem{procedure}{Procedure}

\begin{document}

\title{Braids of entangled particle trajectories}

\author{Jean-Luc Thiffeault}
%\thanks{Accepted for publication in a special issue of \emph{Chaos}
%   on Lagrangian Coherent Structures (2009)}
\email{jeanluc@math.wisc.edu}
\affiliation{Department of Mathematics, University
  of Wisconsin, Madison, WI 53706, USA}

\date{\today}

%\pacs{47.52.+j, 05.45.-a}

\begin{abstract}
  In many applications, the two-dimensional trajectories of fluid
  particles are available, but little is known about the underlying
  flow.  Oceanic floats are a clear example.  To extract quantitative
  information from such data, one can measure single-particle
  dispersion coefficients, but this only uses one trajectory at a
  time, so much of the information on relative motion is lost.  In
  some circumstances the trajectories happen to remain close long
  enough to measure finite-time Lyapunov exponents, but this is rare.
  We propose to use tools from braid theory and the topology of
  surface mappings to approximate the topological entropy of the
  underlying flow.  The procedure uses all the trajectory data and is
  inherently global.  The topological entropy is a measure of the
  entanglement of the trajectories, and converges to zero if they are
  not entangled in a complex manner (for instance, if the trajectories
  are all in a large vortex).  We illustrate the techniques on some
  simple dynamical systems and on float data from the Labrador sea.
  The method could eventually be used to identify Lagrangian coherent
  structures present in the flow.
\end{abstract}

\maketitle

\textbf{
  Consider particles floating on top of a fluid.  We can follow their
  trajectories, either with a camera or by computer simulation.  If we
  then plot their position in a three-dimensional graph, with time the
  vertical coordinate, we get a `spaghetti plot,' which contains
  information about how entangled the trajectories are.  We discuss
  how to measure the level of entanglements in terms of topological
  entropy, and the interpretation of the results.  This provides a
  straightforward method of estimating the level of chaos present in a
  system.  This approach could also be used to determine if some
  trajectories remain together for a long time, and are thus part of a
  Lagrangian coherent structure.
}

\section{Introduction}

\subsection{Floats in the ocean: an example}
\label{sec:intro_floats}

\begin{figure}
\begin{center}
\subfigure[]{
  \includegraphics[width=.42\textwidth]{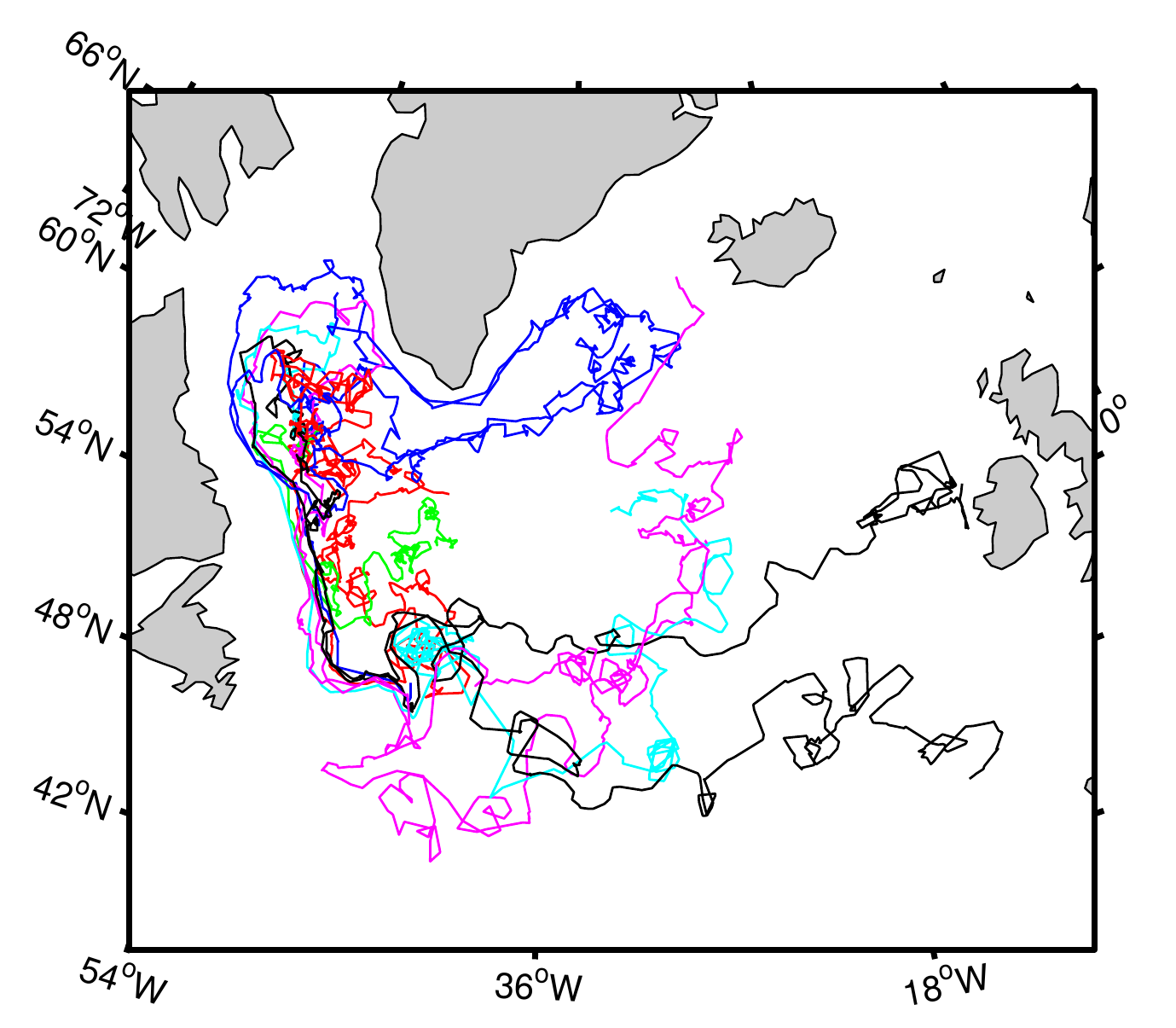}
  \label{fig:floats}
}\hspace{1em}
\subfigure[]{
  \includegraphics[width=.47\textwidth]{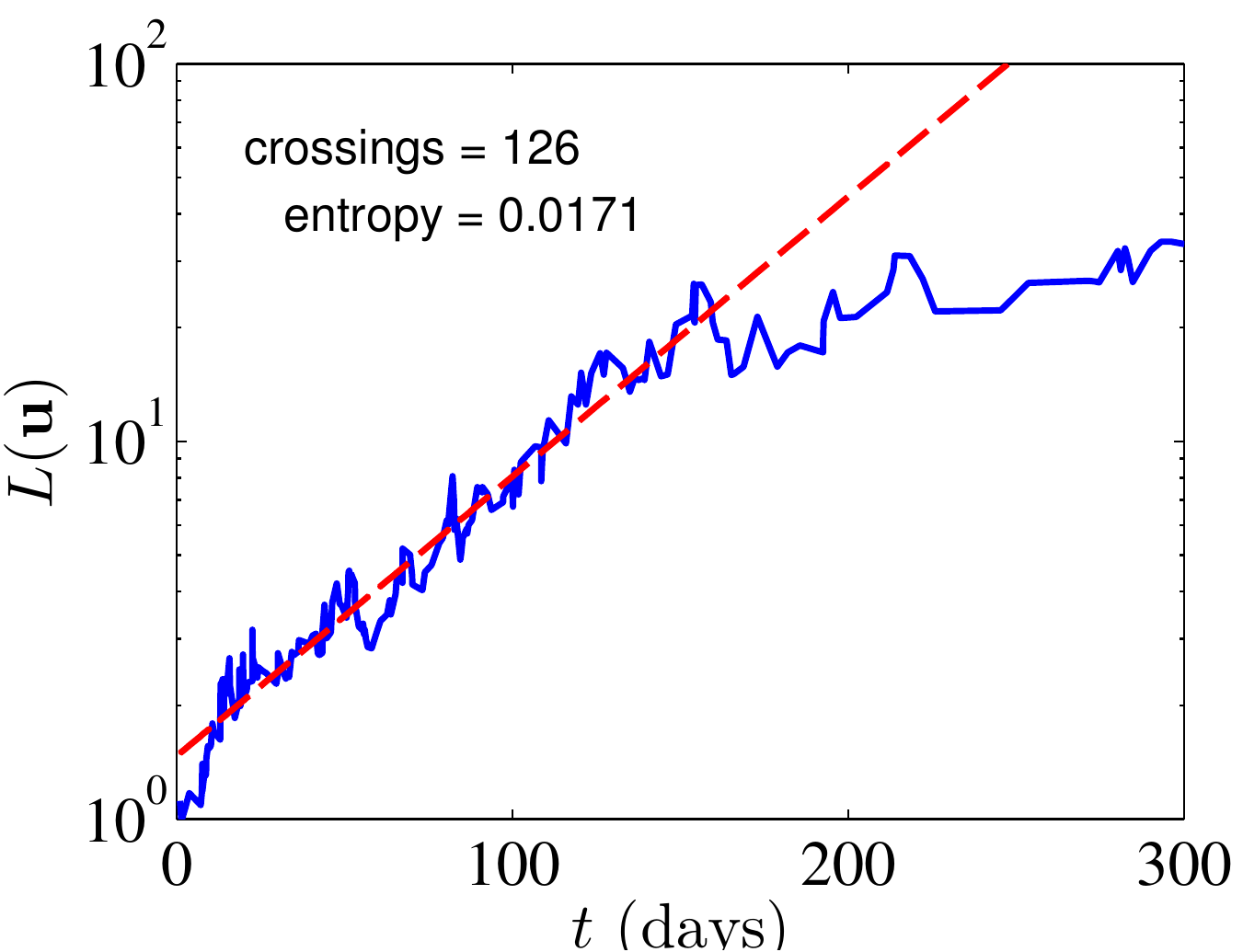}
  \label{fig:floats_braid}
}
\end{center}
\caption{(a) 10 floats from Davis' Labrador sea data~\cite{WFDAC}.
  (b) The growth of~$\Nint(\abv)$ for the 10 floats, with a fit of the
  topological entropy.  For longer times, the floats leave the
  Labrador sea and~$\Nint(\abv)$ becomes constant.  The details of the
  analysis are presented in Section~\ref{sec:random}.}
\label{fig:floats_both}
\end{figure}

Figure~\ref{fig:floats} shows the trajectories of ten floats released
in the Labrador sea, for a period of a few months.
The principal reason to release such floats is the data they measure
and transmit back -- temperature, salinity, pressure, etc.  But the
actual trajectories of the floats are also important, since they tell
us something about large-scale transport in the ocean, a crucial
component in understanding global circulation.  From a single float
one can deduce the single-particle dispersion coefficient, a crude
measure of how quickly a float wanders away from its release point.
However, it is better to measure quantities that involve several
floats~\cite{LacasceAosta2004}.  For instance, if floats happen to
start near each other, then we can see how quickly they separate and
measure finite-time Lyapunov exponents~\cite{Wolf1985}, which are
linked to chaotic advection~\cite{Aref1984}.  But if the floats are
nowhere near each other, then a more global quantity is needed.  In
this paper we propose to examine the `braid' defined by the
trajectories and to measure their degree of entanglement.  (All these
terms will be defined more precisely.)  The number we get out of this
is called the braid's \emph{topological entropy}.
Figure~\ref{fig:floats_braid} shows a measure of the entanglement of
the ten floats as a function of time, with an exponential fit: the
growth rate is the topological entropy.  (For longer times, the curve
levels off because floats start leaving the Labrador sea, which is not
really a closed system.)  Much like a Lyapunov exponent, the
topological entropy gives us a characteristic time for the
entanglement of the floats in the Labrador sea, here about~$1/0.02
\simeq 50$ days.

\subsection{Topology and trajectories}

Since the original paper of Boyland \etal~\cite{Boyland2000},
topological techniques in fluid dynamics have been applied to free
point vortices~\cite{Boyland2003}, fixed blinking
vortex~\cite{Thiffeault2005,Kin2005}, rod stirring
devices~\cite{MattFinn2003,Vikhansky2004,Gouillart2006,
  Binder2008,Thiffeault2008b}, and spatially-periodic
systems~\cite{MattFinn2006,MattFinn2007}.  (See~\cite{Thiffeault2006}
for a brief review.)  More recently, the emphasis has shifted to
locating periodic orbits that play an important role in
stirring~\cite{Gouillart2006,Binder2008} --- so-called \emph{ghost
  rods} --- and even to the manufacture of such
orbits~\cite{Stremler2007}.  Most of these papers study periodic
motions of rods or particle orbits.  For many practical applications,
however, periodic motion is not directly observable, since most such
orbits are highly unstable.  Hence, some authors have examined
\emph{random braids}~\cite{Berger1990,
  Vikhansky2003,Thiffeault2005,Thiffeault2006,MattFinn2007} composed
of arbitrary chaotic trajectories.  (There is also related literature
from the knot theory perspective --- see for example~\cite{Nechaev}.)

The goal of the present paper is to give concrete techniques that can
be used to obtain topological information from particle
trajectories.  The mathematical details are glossed over: the emphasis
is on usability.  Implementation details are discussed, and some
sample Matlab programs are presented in an appendix.  The hope is that
this will make these techniques more accessible to those with little
or no background in braid theory and topology.

The principal measurement we extract from a braid is its
topological entropy.  This entropy is closely related to the
traditional Lyapunov exponent, except that being a topological
quantity it is not sensitive to the size of the sets on which chaos is
occurring.  This is both a weakness and a strength: it does not tell
us everything we might like to know, but on the other hand the
topological entropy is easy to compute from crude data.  This is in
contrast to Lyapunov exponents, which require at the very least
detailed knowledge of particle trajectories that start close together,
and at best the velocity field and its gradient.  When dealing with,
for example, data from oceanic floats (as we will later in this
paper), being able to compute a Lyapunov exponent is a rarity.

There is also a philosophical point that bears some discussion.  The
viewpoint of the present paper is that given particle trajectory data,
a useful thing to quantify is how `entangled' the particle
trajectories are.  This can be done from the particle data directly,
without worrying about the underlying flow.  By contrast, Lyapunov
exponents are defined locally and are sensitive to the \emph{smooth
  structure} of the flow.  It is exactly the (presumed) smooth nature
of the flow that connects local information to a global quantity such
as the Lyapunov exponent, but one is left wondering why we should care
about the local picture at all in practical situations.  The
topological viewpoint presented here is an attempt to sidestep this
and focus directly on global information.

We begin in Section~\ref{sec:braids} by a short introduction to braid
theory, surface dynamics, and their connection to dynamical systems.
In Section~\ref{sec:braidsfromflow} we show how to extract braids from
particle trajectory data.  Section~\ref{sec:entropy} is devoted to
topological entropy: in Section~\ref{sec:entrflow} we discuss its
connection to flows, and in Section~\ref{sec:entrbraid} we show how to
measure it from a braid (for a braid corresponding to periodic
orbits).  In Section~\ref{sec:random} we introduce random braids, and
again show how to measure entropies.  As an application, we calculate
the entropy for floats in the Labrador sea, as presented in
Fig.~\ref{fig:floats_both}.  We offer some concluding remarks in
Section~\ref{sec:discussion}.

\section{Braids}
\label{sec:braids}

\subsection{Physical and Algebraic braids}
\label{sec:braidgroups}

\begin{figure}
\begin{center}
\subfigure[]{
  \includegraphics[height=.15\textheight]{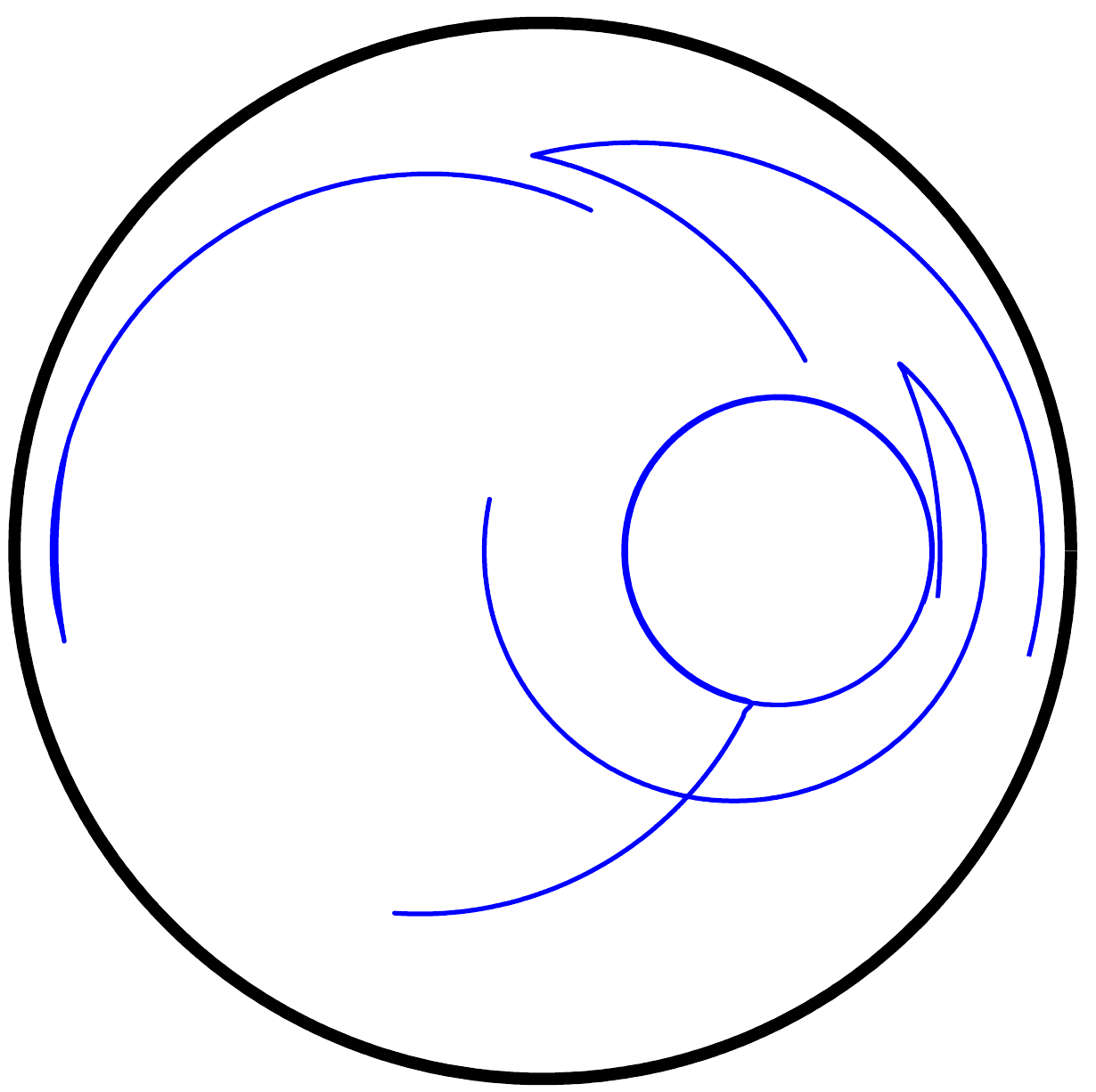}
  \label{fig:orbits}
}\hspace{2em}%
\subfigure[]{
  \includegraphics[height=.3\textheight]{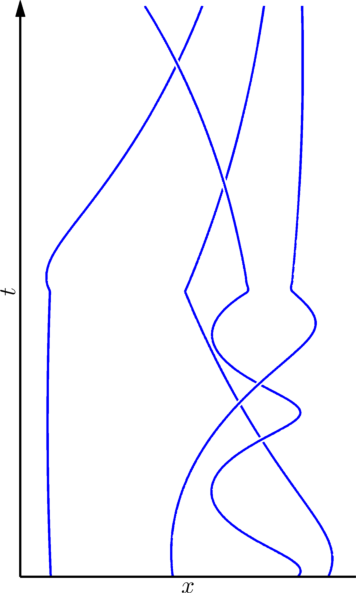}
  \label{fig:orbits_braid}
}\hspace{2em}%
\subfigure[]{
  % -3    -2    -3     2     1
  \includegraphics[height=.3\textheight]{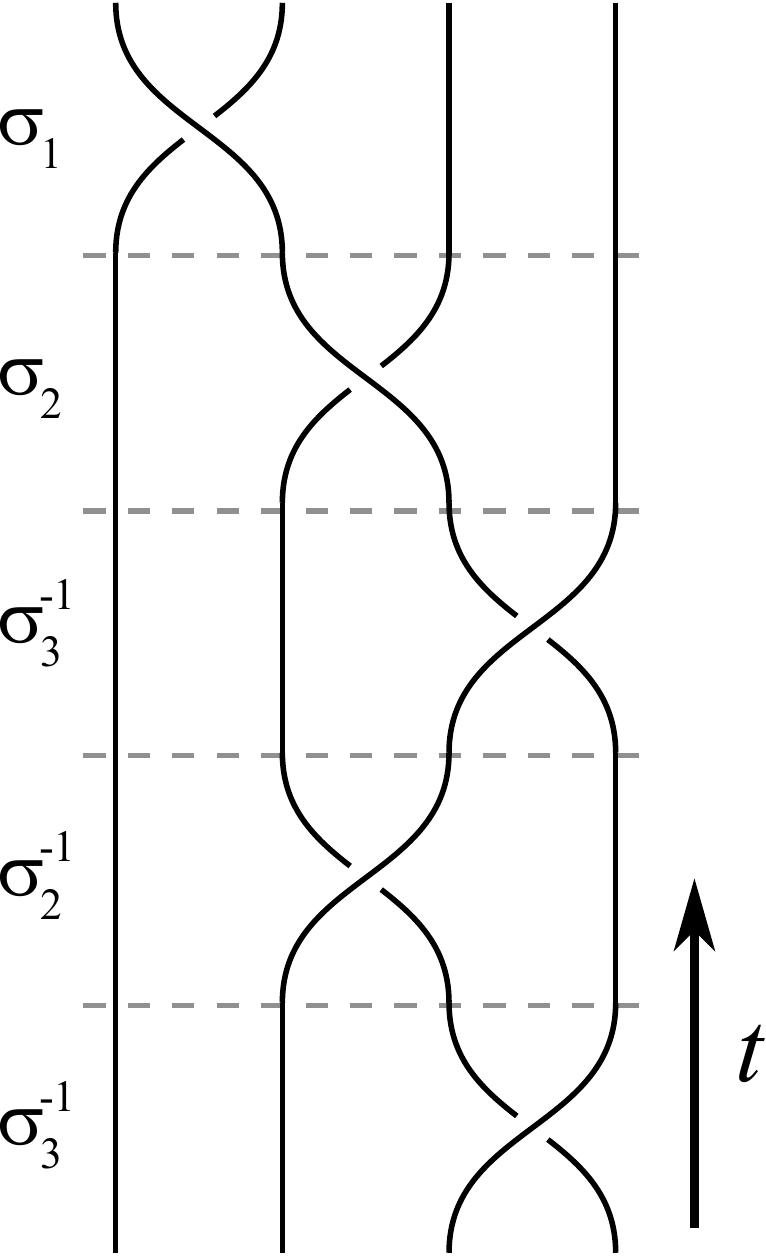}
  \label{fig:braid}
}
\end{center}
\caption{(a) The orbits of four particles in a circular
  two-dimensional domain; (b) The same orbits, lifted to a space-time
  diagram in three dimensions, with time flowing from bottom to top.
  (c) The standard braid diagram corresponding to (b).}
\label{fig:braidexample}
\end{figure}

First we describe intuitively how braids arise.
Figure~\ref{fig:orbits} shows the orbits of 4 particles in a circular
two-dimensional domain.  The particles might be fluid elements,
solutions of ordinary differential equations, or physical particles at
the surface of a fluid.  Figure~\ref{fig:orbits_braid} shows the
`world line' of the same orbits: they are plotted in a three-dimensional
graph, with time flowing vertically upwards.  The diagram in
Fig.~\ref{fig:orbits_braid} depicts a \emph{physical braid}, made up
of four strands.  No strand can go through another strand as a
consequence of the deterministic motion of the particles (they never
occupy the same point at the same time).
Moreover, the mathematical definition of a braid requires that strands
cannot `loop back': here this simply means that the particles cannot
travel back in time.  We will say that two braids are equivalent if they
can be deformed into each other with no strand crossing other strands
or boundaries.  Throughout this paper, we will be interested in
characterizing the level of `entanglement' of trajectories.

\begin{figure}
\begin{center}
\subfigure[]{
  \includegraphics[width=.45\textwidth]{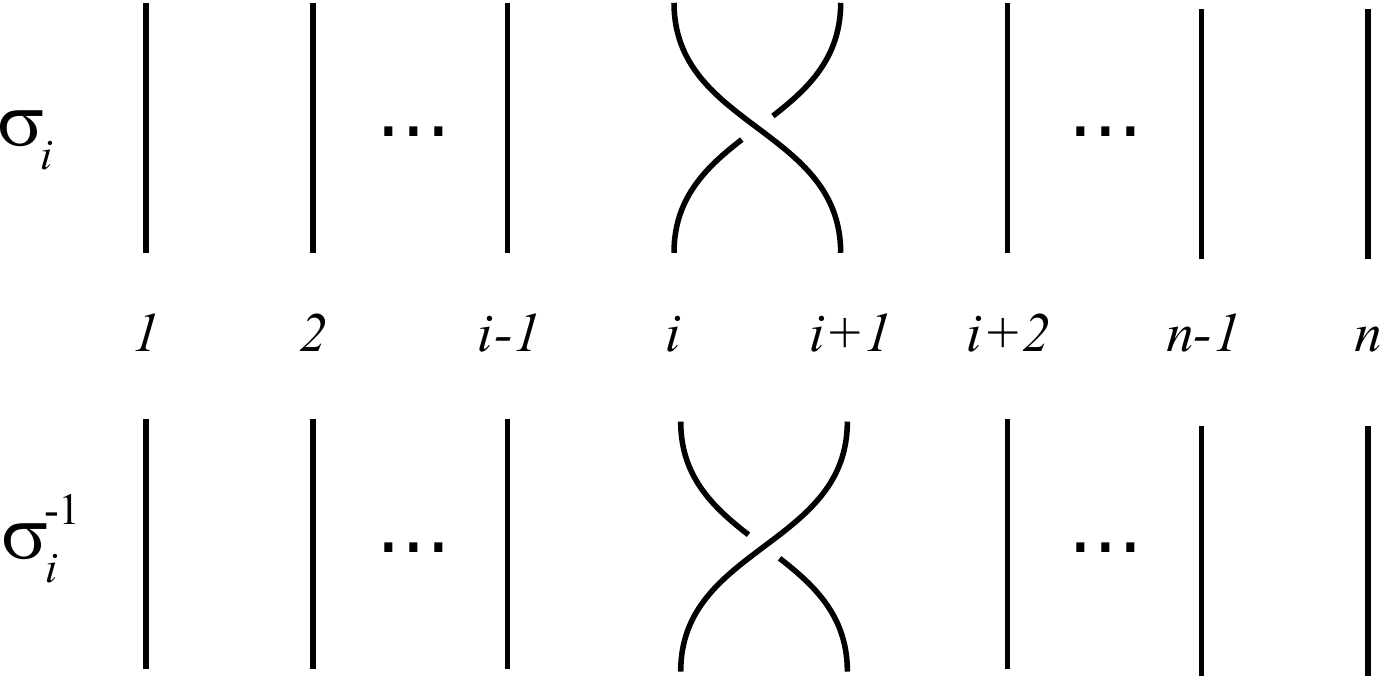}
  \label{fig:sigmas}
}\\
\subfigure[]{
  \includegraphics[width=.55\textwidth]{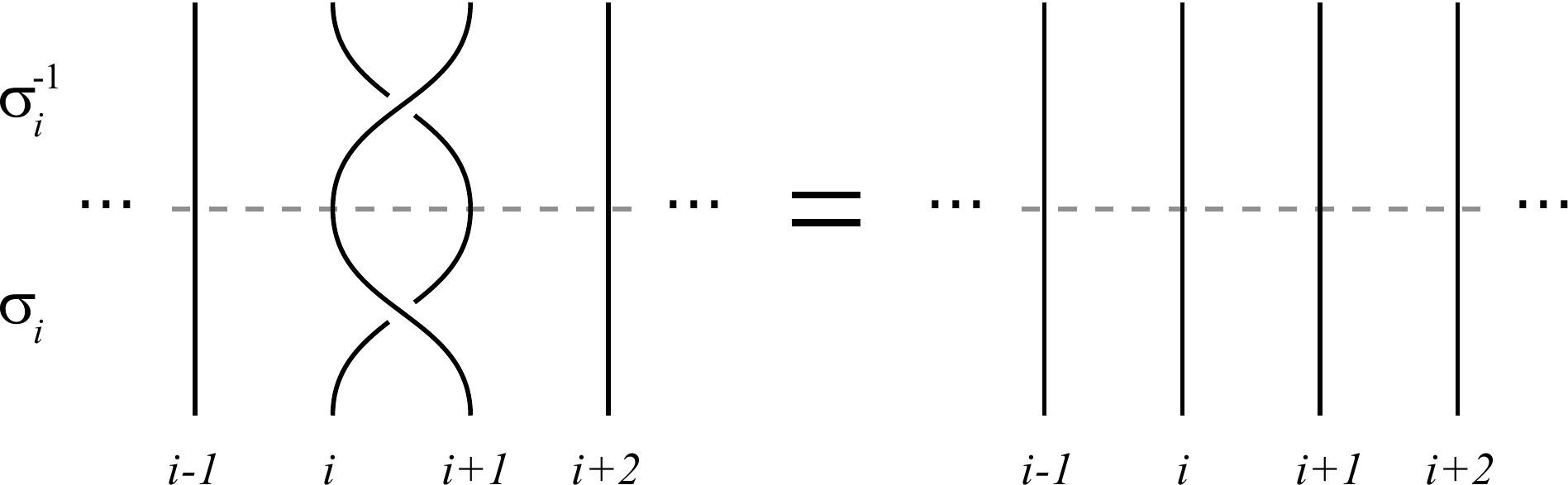}
  \label{fig:inverses}
}
\end{center}
\caption{(a) Braid group generator~$\sigma_\ip$, corresponding to the
  clockwise interchange of the string at the~$\ip$th and~$(\ip+1)$th
  position, counted from left to right.  Its inverse~$\sigma_\ip^{-1}$
  involves their counterclockwise exchange. (b) The concatenation
  of~$\sigma_\ip$ and~$\sigma_\ip^{-1}$ gives the identity braid.}
\end{figure}

Since we can move the strands, is convenient to draw braids in a
normalized form, as shown in Fig.~\ref{fig:braid} for the braid in
Fig.~\ref{fig:orbits_braid}.  Such a picture is called a \emph{braid
  diagram}.  The important thing is that we record when crossings
occur, and which particle was behind and which was in front.  It
matters little how we define `behind,' as long as we are consistent
(see Section~\ref{sec:braidsfromflow} for practical considerations).
In Fig.~\ref{fig:braid} the horizontal dashed lines also suggest that
we can divide the braid into a sequence of elementary crossings, known
as \emph{generators}.  Figure~\ref{fig:sigmas} shows the definition
of~$\sigma_\ip$, which denotes the clockwise interchange of the
$\ip$th and $(\ip+1)$th strands, keeping all other strands fixed.
Note that the index~$\ip$ is the position of the strand from left to
right, \emph{not} a label for the particular strand.  For~$\nn$
strands, we have~$\nn-1$ distinct generators.

Figure~\ref{fig:sigmas} also shows the counterclockwise interchange of
two strands, denoted by the operation~$\sigma_\ip^{-1}$.  The
justification of the `inverse' notation is evident in
Fig.~\ref{fig:inverses}: if we concatenate~$\sigma_\ip$
and~$\sigma_\ip^{-1}$, then after pulling tight on the strands we find
that they are disentangled.  We call the braid on the right in
Fig.~\ref{fig:inverses} the \emph{identity braid}.  In fact, the set
of all braids on a given number~$\nn$ of strands forms a \emph{group}
in the mathematical sense: the group operation is given by
concatenation of strands, the inverse by reversing the order and
direction of crossings, the identity is as described above, and it is
clear that concatenation is associative.  This group is
called~$\Br_\nn$, the braid group on~$\nn$ strands, also known as the
Artin braid group.

The braid group~$\Br_\nn$ is generated by the
set~$\{\sigma_1,\ldots,\sigma_{\nn-1}\}$: this means that any braid
in~$\Br_\nn$ can be written as a product (concatenation)
of~$\sigma_\ip$'s and their inverses.  The braid group is
\emph{finitely-generated}, even though it is itself infinite: only a
finite number of generators give the whole group.  To see that the
braid group contains an infinite number of braids, simply
consider~$\sigma_1^k$, for~$k$ an arbitrary integer: no matter how
large~$k$ gets, we always get a new braid out of this, consisting of
increasingly twisted first and second strands.

\begin{figure}
\begin{center}
\subfigure[]{
  \includegraphics[width=.75\textwidth]{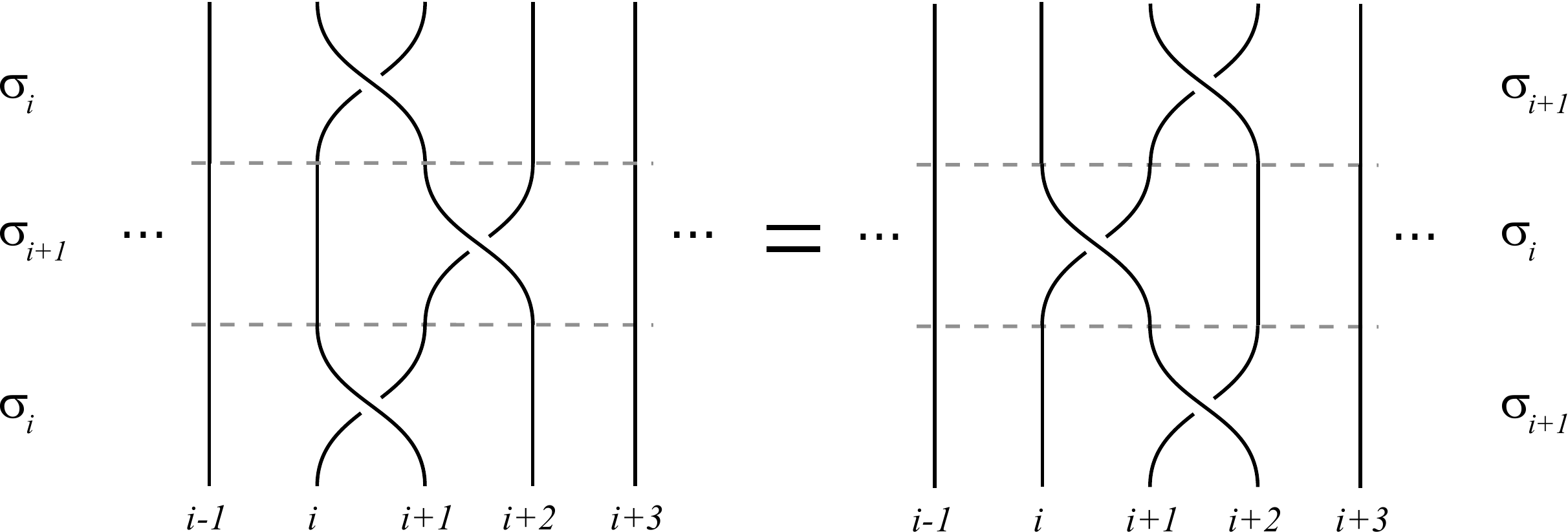}
  \label{fig:relation1}
}\\
\subfigure[]{
  \includegraphics[width=.65\textwidth]{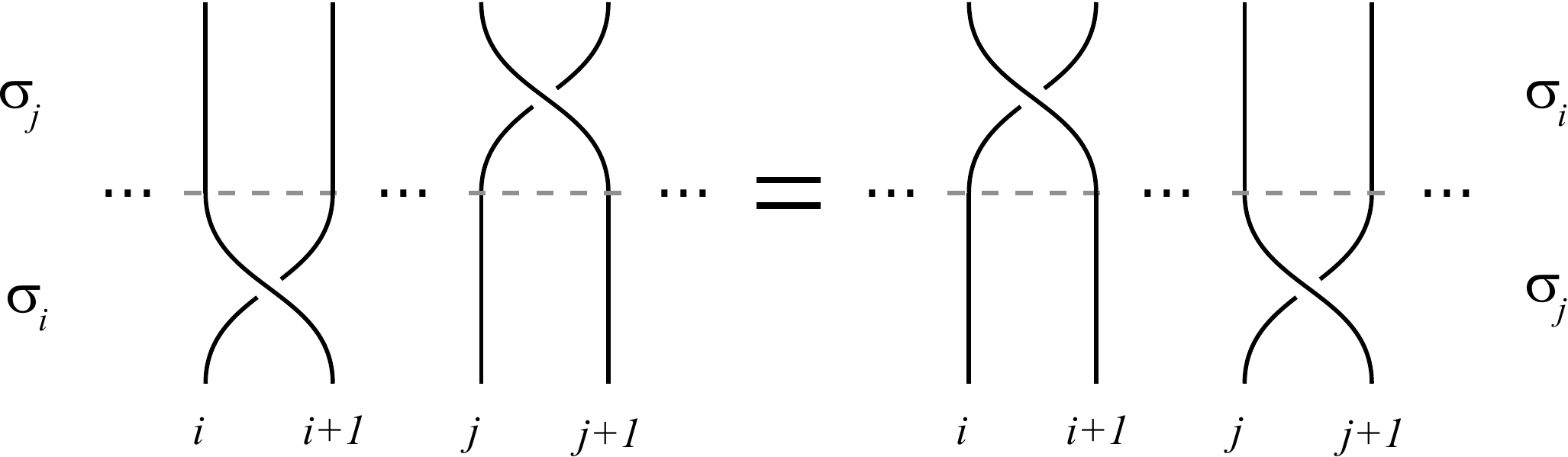}
  \label{fig:relation2}
}
\end{center}
\caption{Braid group relations (see Eq.~\eqref{eq:presentation}): (a)
  relation for three adjacent strands; (b) commutation relation for
  generators that don't share a strand.}
\end{figure}

We have now passed from physical braids, as depicted in
Fig.~\ref{fig:orbits_braid}, to \emph{algebraic braids}.  The
algebraic braid corresponding to Fig.~\ref{fig:braid}
is~$\sigma_3^{-1}\sigma_2^{-1}\sigma_3^{-1}\sigma_2\sigma_1$,
where we read generators from left to right in time (beware:
conventions differ).  In essence, an algebraic braid is simply a
sequence of generators, which may or may not come from a physical
braid.  How can we guarantee that physical braids and algebraic braids
describe the same group?  We need to be mindful of \emph{relations}
amongst the generators that arise because of physical constraints.
For example, Fig.~\ref{fig:relation1} shows a relation amongst
adjacent triplets of strands.
Staring at the picture long enough, and allowing for the deformation
of strands without crossing, the reader can perhaps see the that
braids in Fig.~\ref{fig:relation1} are indeed equal.  Hence, the
algebraic sequence~$\sigma_\ip\sigma_{\ip+1}\sigma_\ip$ must be equal
to~$\sigma_{\ip+1}\sigma_\ip\sigma_{\ip+1}$, if the generators are to
correspond to physical braids.  Another, more intuitive relation is
shown in Fig.~\ref{fig:relation2}: generators commute if they do not
share a strand.  In summary, we have the relations
\begin{subequations}
\begin{alignat}{2}
\sigma_\ip \sigma_\jp \sigma_\ip &= \sigma_\jp \sigma_\ip \sigma_\jp \qquad
&\text{if } |\ip-\jp| &= 1, \label{eq:121} \\
\sigma_\ip \sigma_\jp &= \sigma_\jp \sigma_\ip \qquad &\text{if }
|i-j| &> 1,
\label{eq:comm}
\end{alignat}%
\label{eq:presentation}%
\end{subequations}%
amongst the generators.  Artin~\cite{Artin1947} proved the surprising
fact that there are no other relations satisfied by the
generators~$\sigma_\ip$, except for those than can be derived
from~\eqref{eq:presentation} by basic group operations
(multiplication, inversion, etc.).  The
generators~$\{\sigma_1,\ldots,\sigma_{\nn-1}\}$ together with the
relations~\eqref{eq:presentation} define the \emph{algebraic braid
  group}, which we also denote~$\Br_\nn$.  With these relations, the
groups of physical and algebraic braids are isomorphic.

A consequence of the relations~\eqref{eq:presentation} is that it may
not be immediately obvious that two algebraic braids are equal.  For
instance, the braids~$\sigma_1\sigma_2$
and~$\sigma_2\sigma_2\sigma_1\sigma_2\sigma_1^{-1}\sigma_1^{-1}$ are
equal, since
\begin{equation*}
  \sigma_1\sigma_2 =
  (\sigma_1\sigma_2\sigma_1)\sigma_1^{-1} =
  (\sigma_2\sigma_1\sigma_2)\sigma_1^{-1} =
  \sigma_2(\sigma_1\sigma_2\sigma_1)\sigma_1^{-1}\sigma_1^{-1} =
  \sigma_2\sigma_2\sigma_1\sigma_2\sigma_1^{-1}\sigma_1^{-1}\,.
\end{equation*}
This `braid equality' problem has seen many refinements: the original
solution of Artin~\cite{Artin1947} has computational complexity
exponential in the number of generators, but modern techniques can
determine equality in a time quadratic in the braid
length~\cite{Birman1998,Birman2004,Dynnikov2007}.

\subsection{Extracting the braid from a flow}
\label{sec:braidsfromflow}

The first step in obtaining useful topological information from
particle trajectories is to compute their associated braid,
essentially going from the physical picture in
Fig.~\ref{fig:orbits_braid} to the algebraic picture in
Fig.~\ref{fig:braid}.  A simple method to do this was originally
described in~\cite{Thiffeault2005}, but is also implicit in earlier
work such as~\cite{Gambaudo1999,Vikhansky2003} (see
also~\cite{Lefranc2006} for a related technique).

We start with trajectory information for~$\nn$ particles over some
time.  We first project the position of the particles onto any fixed
\emph{projection line} (which we choose to be the horizontal axis),
and label the particles by~$\ip=1,2,\ldots,\nn$ in increasing order of
their projection.  A crossing occurs whenever two particles
interchange position on the projection line.  A crossing can occur as
an ``over'' or ``under'' braid, which for us means a clockwise or
counterclockwise interchange.  These interchanges correspond to the
braid group generators introduced in Section~\ref{sec:braidgroups}.

\begin{figure}
\begin{center}
\subfigure[]{
  \includegraphics[height=.35\textwidth]{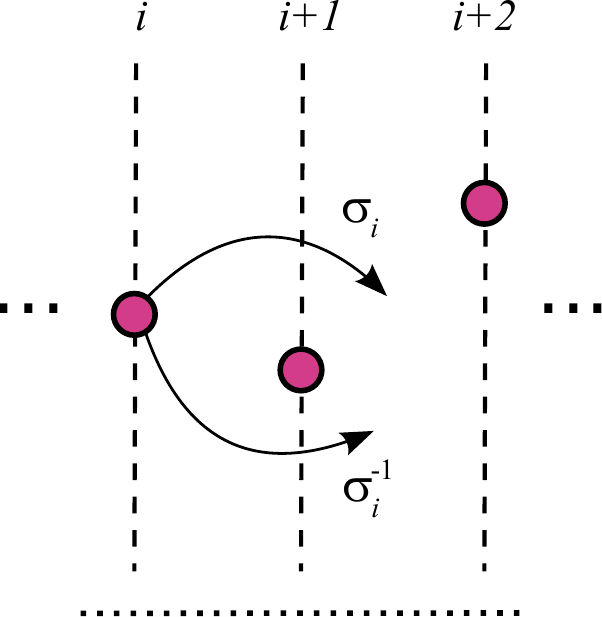}
  \label{fig:crossinga}
}\hspace{4em}%
\subfigure[]{
  \includegraphics[height=.35\textwidth]{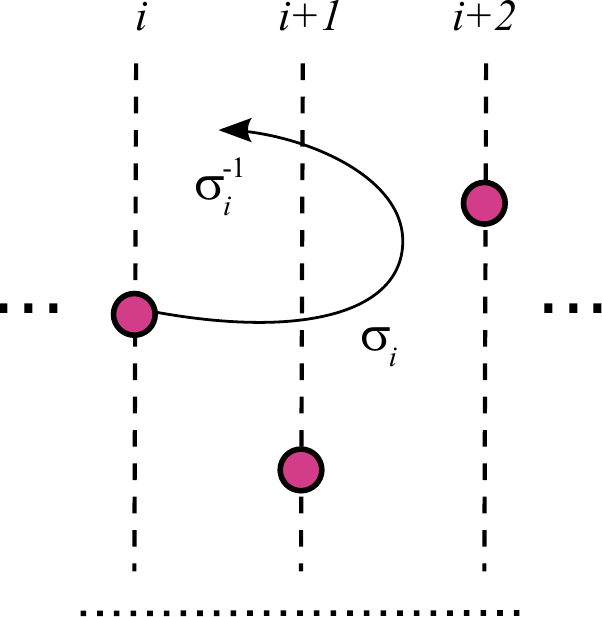}
  \label{fig:crossingb}
}
\end{center}
\caption{Detecting crossings: (a) Two possible particle paths that are
  associated with different braid group generators; (b) Two crossings
  that yield no net braiding.  The projection line used to detect
  crossings is shown dotted at the bottom, and the perpendicular lines
  used to determine the braid generator are shown dashed.}
\label{fig:crossing}
\end{figure}

Assuming a crossing has occurred between the $\ip$th and $(\ip+1)$th
particles, we need to determine if the corresponding braid generator
is~$\sigma_\ip$ or~$\sigma_\ip^{-1}$.  We look at the projection of
the $\ip$th and~$(\ip+1)$th particles in the direction perpendicular
to the projection line (the vertical axis in our case).  If the $\ip$th
particle is \emph{above} the~$(\ip+1)$th at the time of crossing, then
the interchange involves the group generator~$\sigma_\ip$ (we define
``above'' as having a greater value of projection along the
perpendicular direction).  Conversely, if the $\ip$th particle is
\emph{below} the~$(\ip+1)$th at the time of crossing, then the
interchange involves the group generator~$\sigma_\ip^{-1}$.
Figure~\ref{fig:crossinga} depicts these two situations.

The net result is that from the~$\nn$ particle trajectories we obtain
a time-ordered sequence of the generators~$\sigma_\ip$
and~$\sigma_\ip^{-1}$, $\ip=1,\ldots,\nn-1$.  We call this sequence
the braid of the trajectories.  We also record the times at which
crossings occur, so each generator in the sequence has a time
associated with it.

\bigskip

\noindent\textbf{Remarks:}
\begin{itemize}
\item The method just described might seem to detect spurious
  crossings if two well-separated particles just happen to interchange
  position on the projection line several consecutive times in a row,
  as shown in Figure~\ref{fig:crossingb}.  However, this would imply
  a sequence of~$\sigma_\ip$ and~$\sigma_\ip^{-1}$ braid generators,
  since which particle is the~$\ip$th one changes at each crossing.
  When composed together the generators for these crossings cancel.
\item We give a simple Matlab implementation of the method in
  Appendix~\ref{code:gencross}.  The program~\texttt{gencross} detects
  crossings in trajectory data; it makes an effort to resolve multiple
  simultaneous crossings (up to triple crossings), but will complain
  if it gets confused.  More sophisticated code can be written that
  re-interpolates the trajectory as needed to detect crossings.
\item When the system has symmetries, such as when several periodic
  orbits lie on the same line, there are `bad' choices of projection
  line where it is impossible to resolve the order of crossings, since
  orbits cross at exactly the same time.  Displacing the projection
  line a little cures this.
\item If the braid is truly periodic, that is, if all the particles
  return to their initial configuration, then changing the projection
  line changes the braid, but only by \emph{conjugation}, which does
  not affect the entropy~\cite{Boyland2003}
  (Section~\ref{sec:entropy}).
\item If the trajectories are not periodic, then the method does not
  define a braid in the traditional sense where all the strands return
  to the same initial configuration.  This is inconsequential to our
  purposes: all that matters is the order along the projection line
  (see also~\cite{Gambaudo1999}).  The choice of projection line
  changes the braid beyond simple conjugation, but this only creates
  an error in a small, finite number of generators, which is not
  important when considering long braids and does not asymptotically
  affect the entropy (Section~\ref{sec:entropy}).
\item If the braid is generated from chaotic trajectories, then
  missing a few crossings (due to, say, gaps in the data) is fine as
  long as the trajectories are long enough.
\end{itemize}

\section{Topological entropy}
\label{sec:entropy}

In Section \ref{sec:braids} we described how a set of trajectories in
the two-dimensional dynamical system can be described as a braid in a
three-dimensional space-time diagram.  In this section we will
describe further how this braid relates to topological information for
the underlying flow.

It is worth noting that braids are not always interpreted in terms of
trajectories: they arose first and are still studied as independent
geometrical and algebraic objects.  The reason they take center stage
in the present study is through their connection to mappings of
surfaces (mapping class groups).  The Thurston--Nielsen
theory~\cite{Birman1975,Thurston1988,Fathi1979,Casson1988,Boyland1994}
classifies mapping of surfaces according to whether they can be
``deformed'' to each other in a topological sense.  Braids provide a
convenient way of labeling the \emph{isotopy classes} that result.  So
even though we will often speak here of the braid as being the primary
object of interest, we are really using techniques that apply to the
class of mappings labeled by a braid.

\subsection{Entropy of a flow}
\label{sec:entrflow}

Ultimately, we want to measure the topological entropy of a system
directly from a braid of trajectories.  Before we do this, we discuss
the meaning of topological entropy of a flow or map.
The topological entropy of a dynamical system measures the loss
of information under the dynamics.  It is closely related to the
Lyapunov exponent, which measures the time-asymptotic rate of
separation of neighboring trajectories.  But it is in some sense a
cruder quantity, since it does not require a notion of distance.  A
positive entropy is associated with chaos, though it tells us nothing
about the size of the chaotic region.  The topological entropy is an
upper bound on the largest Lyapunov exponent of a flow.  The two
are equal only for very simple systems where stretching is uniform.

Though there are more fundamental ways to define it, we shall take our
working definition of entropy to be the \emph{asymptotic growth rate
  of material lines}~\cite{Newhouse1993}.  It is fairly
straightforward to measure this numerically, given a sufficiently
accurate velocity field.  We simply choose an initial material line
and follow it for some time, interpolating new
\begin{figure}
  \includegraphics[width=.6\textwidth]{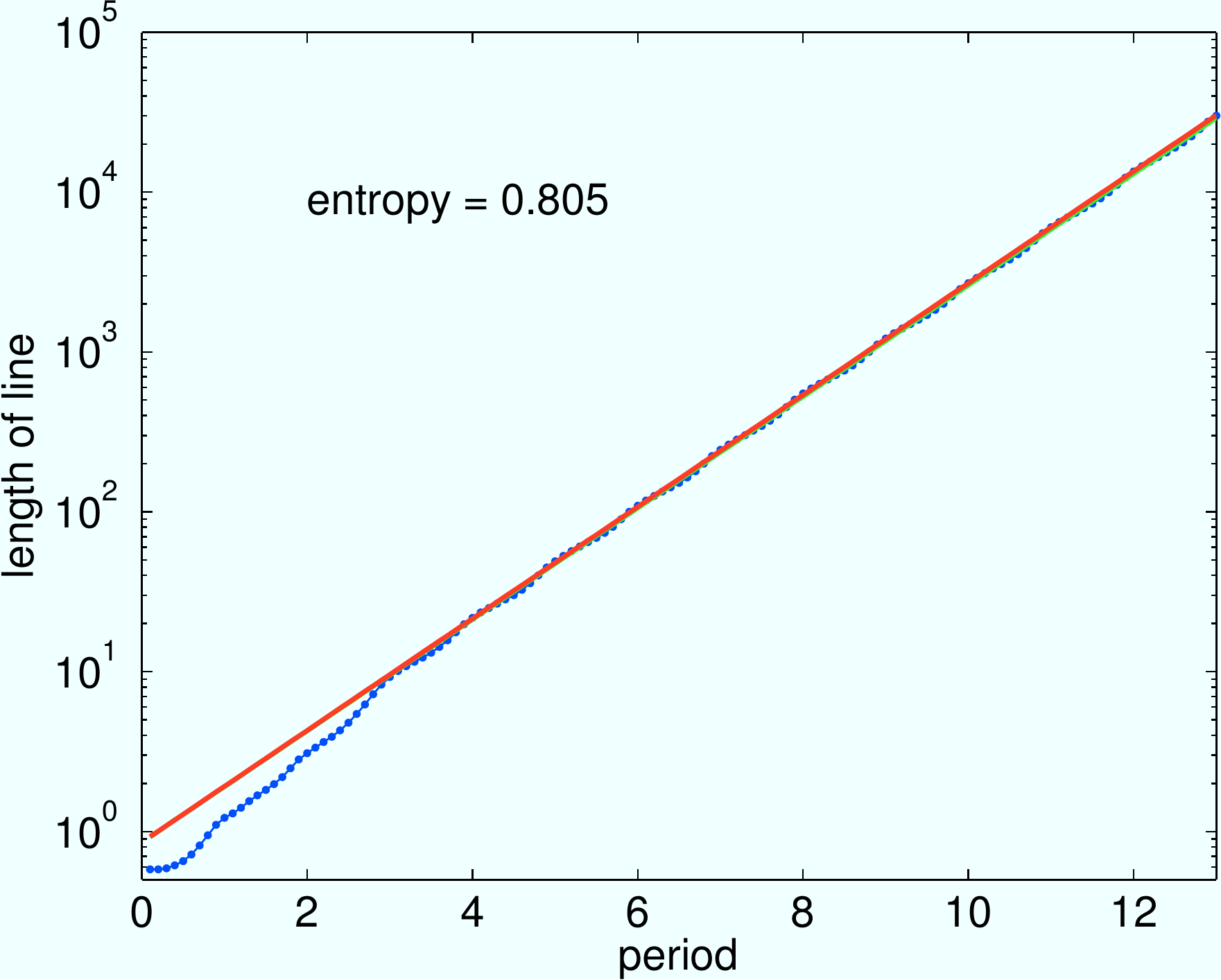}
\caption{The length of a material line as it is advected by a flow for
  many periods.  The exponential growth rate is very well defined
  (fitted line), even though there are identical small oscillations at
  each period.}
\label{fig:linestretch}
\end{figure}
points as the line gets longer.  Figure~\ref{fig:linestretch} shows
such a line for a numerical simulation of a stirred viscous flow.
Note how the exponential growth rate is very sharply defined.  The
topological entropy is the supremum of the growth rate over all such
loops, but in practice almost any nontrivial loop (\ie, that spans the
domain) will grow exponentially at a rate~$\htopf$.

In practical applications we often do not have access to an accurate
representation of the velocity field.  This is where braids come in,
as a way of approximating the topological entropy.  As we will see in
Section~\ref{sec:entrbraid}, the braid provides a lower bound on the
flow's topological entropy.

\subsection{Entropy of a braid}
\label{sec:entrbraid}

\begin{figure}
\begin{center}
  \includegraphics[width=\textwidth]{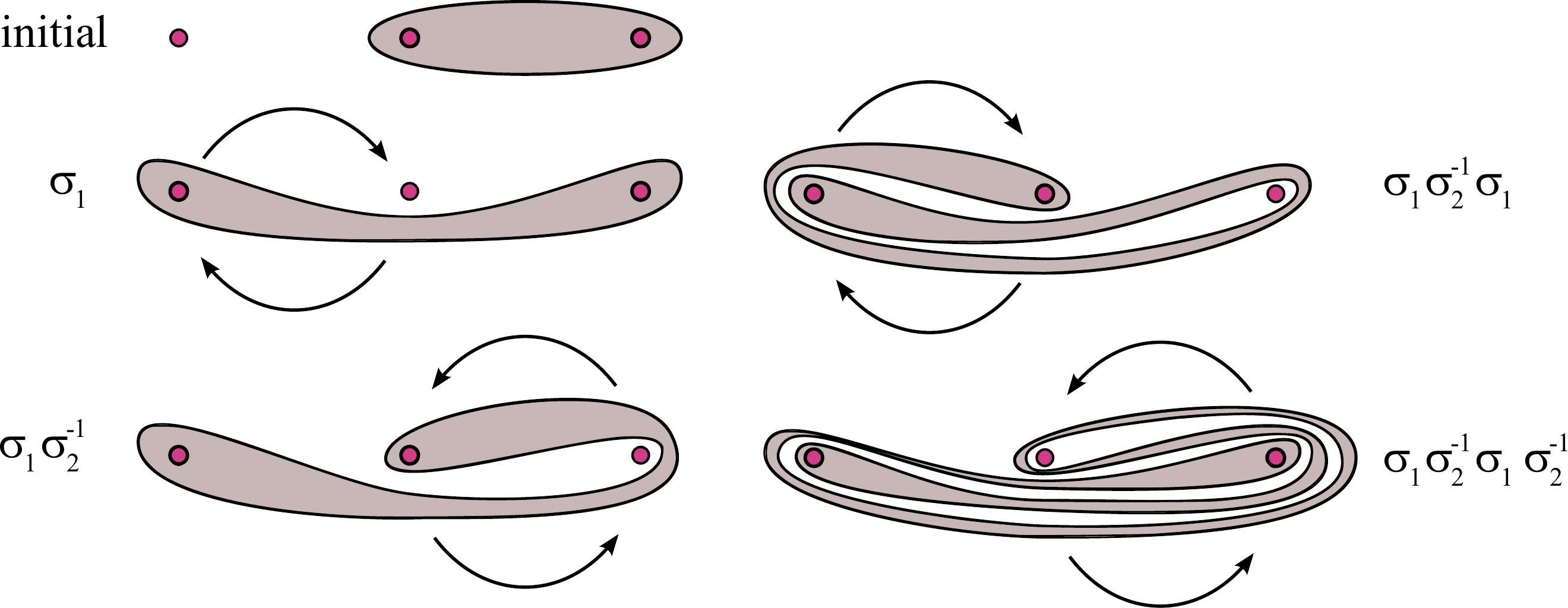}
\end{center}
\caption{For~$\nn=3$ particles, a loop is initially wrapped around the
  second and third particles.  The generator~$\sigma_1$ is applied,
  interchanging clockwise the first and second particles, followed
  by~$\sigma_2^{-1}$, which interchanges the second and third
  particles counterclockwise.  The loop is forced to stretch under this
  application.  The final loop on the lower right underwent two
  applications of the braid~$\sigma_1\sigma_2^{-1}$.  If we keep
  applying this braid, the length of the loop will grow exponentially.}
\label{fig:s1s-2_loop}
\end{figure}

Figure~\ref{fig:s1s-2_loop} illustrates how the motion of~$\nn=3$
point particles can be used to put a lower bound on the topological
entropy, defined here as the growth rate of material lines or loops.
Here, the point particles undergo a motion described by the
braid~$\sigma_1\sigma_2^{-1}$.  Two full periods are shown.  Notice
that an initial loop that is `caught' on the particles is forced to
follow along, since determinism implies that it cannot occupy the same
point in phase space as the particles.  In fact, a straightforward
calculation~\cite{Boyland2000} shows that for this braid the total
length of the loop must grow exponentially at least at a
rate~$\log((3+\sqrt{5})/2)$ per period.  We call this rate the
\emph{topological entropy of the braid}, $\htopb{\nn}$, to distinguish it
from the true topological entropy of the flow, $\htopf$, as defined in
Section~\ref{sec:entrflow}.  We have
\begin{equation}
  \htopf \ge \htopb{\nn}
\end{equation}
for any braid obtained from the motion of~$\nn$ particles in the flow.
Typically, the more particles are included in the braid, the
closer~$\htopb{\nn}$ is to~$\htopf$~\cite{MattFinn2007}.  Note
that~$\htopb{1}$ and~$\htopb{2}$ are always zero.

An essential property of~$\htopb{\nn}$ is that the growth \emph{rate} of
the loop is independent of specific details: for instance, if the
particles are not equally spaced, or if the loop is `tightened' around
the particles, then the length will change, but the asymptotic growth
rate will not, because all these changes amount to an additive
constant in the logarithm, which gets divided by a large time.

We are now faced with a task: given a sequence of
generators~$\sigma_\ip$, measured in some way or obtained numerically
from a flow, what is~$\htopb{\nn}$?  The method used
in~\cite{Thiffeault2005}, based on a matrix representation of the
braid group, only provides a lower bound on the braid entropy.  An
accurate and efficient computation has since become a lot simpler due
to a new algorithm by Moussafir~\cite{Moussafir2006}, who uses a set
of coordinates to encode a loop.  We describe this briefly below; for
more details see~\cite{Dynnikov2002,Moussafir2006,Hall2009}.  The
reader who is mostly interested in using the method can skip to the
end of the section to Procedure~\ref{proc:1}.

\begin{figure}
  \includegraphics[width=.4\textwidth]{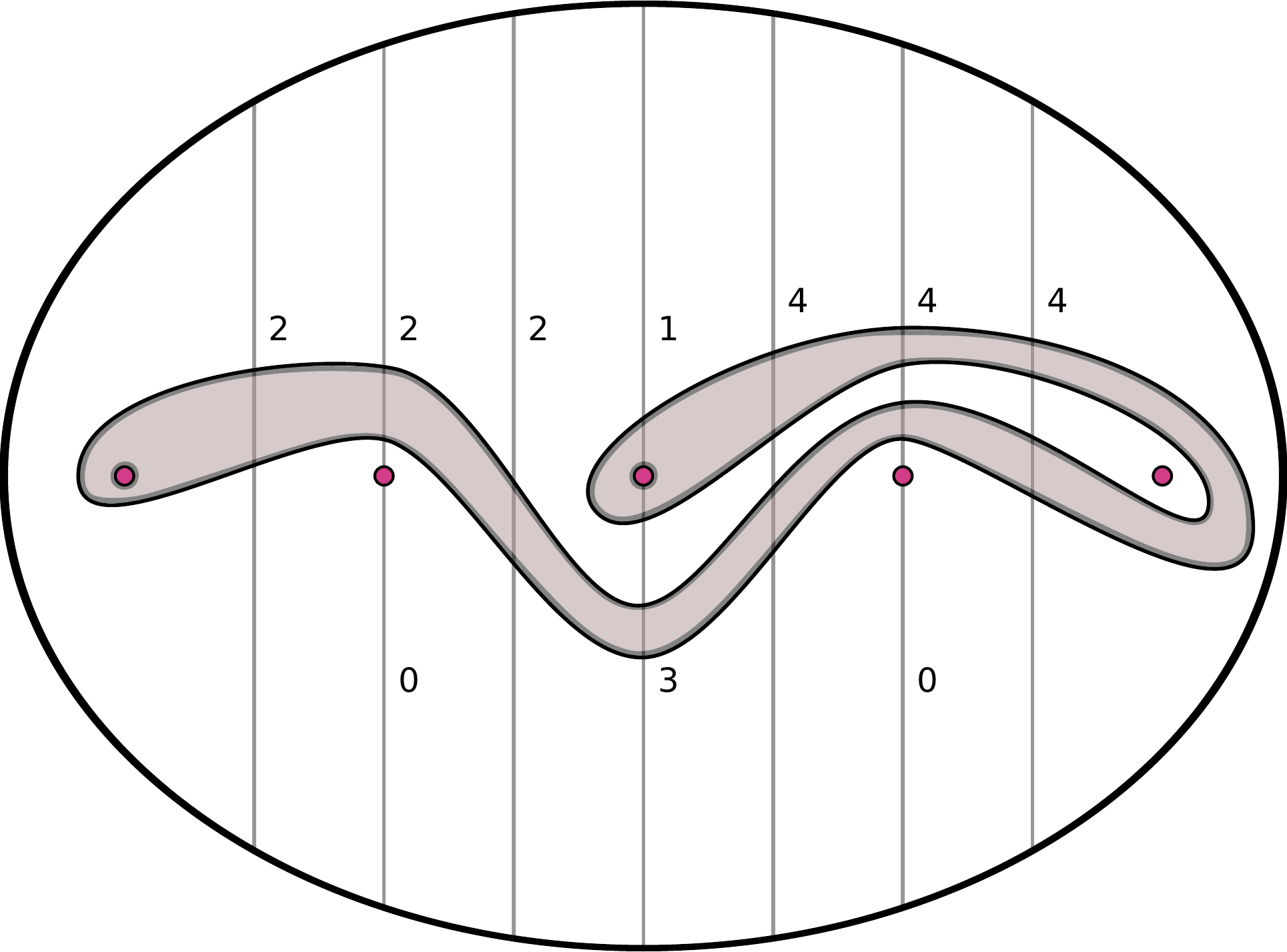}
\caption{A non-intersecting closed loop wrapped around~$\nn=5$
  particles.  Up to trivial deformations, the loop can be
  reconstructed by counting intersections with the vertical lines.  In
  terms of the crossing numbers defined in Fig.~\ref{fig:dynn_def}, this
  loop has~$\nu_1=\nu_2=2$, $\nu_3=\nu_4=4$, $\mu_1=2$, $\mu_3=1$,
  $\mu_5=4$, $\mu_2=\mu_6=0$, $\mu_4=3$, and Dynnikov coordinate
  vector~$\abv=(-1,1,-2,0,-1,0)$ (see Eq.~\eqref{eq:abvdef}).}
\label{fig:dynn_loop}
\end{figure}

The basic idea is simple: consider the closed loop in
Fig.~\ref{fig:dynn_loop}, which is wrapped around~$\nn=5$ particles.
The loop does not intersect itself, so in two dimensions the allowable
paths it can follow around the particles are far from arbitrary.  The
amazing fact is that we can reconstruct the entire loop, or at least
the way it is threaded around the particles, by counting how many
times it intersects the vertical lines in Fig.~\ref{fig:dynn_loop}.

\begin{figure}
  \includegraphics[width=.8\textwidth]{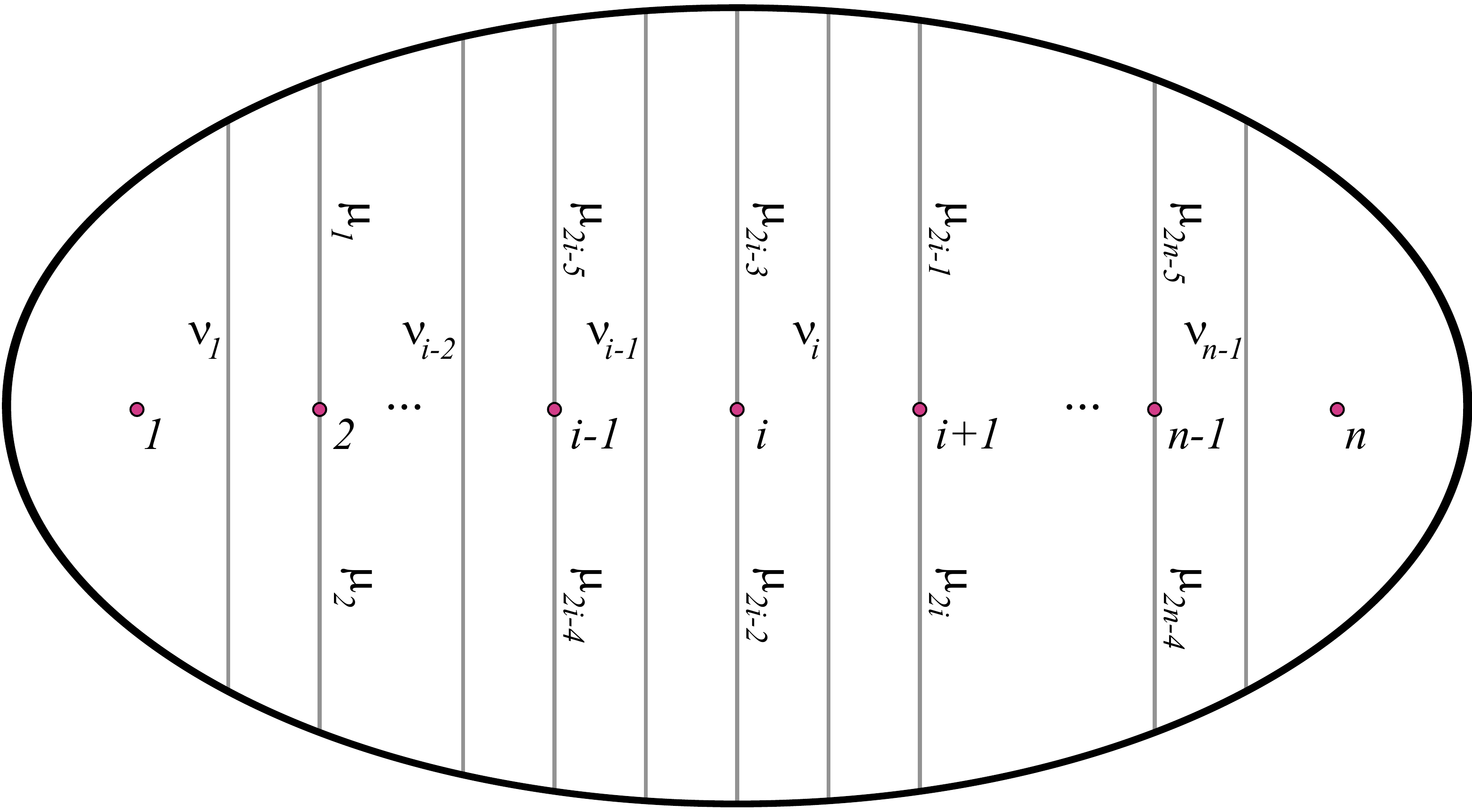}
\caption{Definition of the crossing numbers~$\mu_\ip$ and~$\nu_\ip$.
  The~$\mu_\ip$ for~$\ip$ odd count crossings above a particle, and
  below a particle for~$\ip$ even.  The~$\nu_\ip$ count crossings
  between particles.}
\label{fig:dynn_def}
\end{figure}

In Fig.~\ref{fig:dynn_def} we give specific labels to the
\emph{crossing numbers}.  For~$\nn$ particles, $\mu_{2\ip-3}$ (odd
index) gives the number of crossings of a loop above the~$\ip$th
particle, and~$\mu_{2\ip-2}$ (even index) below the same particle.
The number~$\nu_\ip$ counts the crossings between particle~$\ip$
and~$\ip+1$.  We have a total of~$3\nn-5$ crossing numbers.

This set of crossing numbers (which are all non-negative) can be
reduced further: define
\begin{equation}
  \ac_\ip = \tfrac12\l(\mu_{2\ip} - \mu_{2\ip-1}\r), \qquad
  \bc_\ip = \tfrac12\l(\nu_\ip - \nu_{\ip+1}\r)
\end{equation}
for~$\ip=1,\ldots,\nn-2$.  The vector of length~$(2\nn-4)$,
\begin{equation}
  \abv = (\ac_1,\ldots,\ac_{\nn-2},\bc_1,\ldots,\bc_{\nn-2})
  \label{eq:abvdef}
\end{equation}
is called the \emph{Dynnikov coordinates} of a loop.  (As far as we
know, this specific encoding was originally introduced by
Dynnikov~\cite{Dynnikov2002}, but it is implicit in earlier work of
Thurston, Dehn, and others.)  The components~$\ac_\ip$ and~$\bc_\ip$
are signed integers.  They can be used to exactly reconstruct the
loop~\cite{Hall2009}, but we shall not need to do this here.  This set
of coordinates is minimal: it is not possible to achieve the same
reconstruction with fewer numbers.

We can also obtain the minimum number of intersections~$\Nint(\abv)$
of the loop with the horizontal line through the
particles~\cite{Moussafir2006}:
\begin{equation}
  \Nint(\abv) = \lvert\ac_1\rvert + \lvert\ac_{\nn-2}\rvert
  + \sum_{\ip=1}^{\nn-3}\lvert\ac_{\ip+1}-\ac_\ip\rvert
  + \sum_{\ip=0}^{\nn-1}\lvert\bc_\ip\rvert\,,
  \label{eq:Nint}
\end{equation}
where~$\bc_0$ and~$\bc_{\nn-1}$ can be obtained from the other
coordinates as~\cite{Hall2009}
\begin{equation}
  \bc_0 = -\max_{1\le\ip\le\nn-2} \biggl(\lvert\ac_\ip\rvert
  + \pos{\bc_\ip} + \sum_{\jp=1}^{\ip-1}\bc_\jp\biggr),
  \qquad
  \bc_{\nn-1} = -\bc_0 - \sum_{\jp=1}^{\nn-2}\bc_\ip\,.
\end{equation}
The formula for~$\Nint(\abv)$ is encoded in the Matlab function
\texttt{loopinter} in Appendix~\ref{code:loopsigma}.  For example, the
loop in Fig.~\ref{fig:dynn_loop} intersects the horizontal axis (the
line through all the particles) 12 times.  The crucial observation,
which will allow a simple computation of~$\htopb{\nn}$, is that if the
length of the loop grow exponentially, then~$\Nint(\abv)$ also grows
exponentially at the same rate~\cite{Moussafir2006,Fathi1979}.

Now that we have a way of encoding any loop, we need to find how the
loop is transformed by a braid.  What makes all this work is that
there is a very efficient way of doing this: given a loop encoded
by~$\abv$ as in Eq.~\eqref{eq:abvdef}, each generator of the braid
group~$\sigma_\ip$ simply transforms these coordinates in a
predetermined manner.  (Mathematically, this defines an \emph{action}
of the braid group on the set of Dynnikov coordinates.)  We call these
transformations the \emph{update rules} for a generator.

The update rules are straightforward to code on a computer.  (See
Appendix~\ref{code:loopsigma} for a Matlab implementation.)  To
express them succinctly,\footnote{We are using the numbering scheme
  of~\cite{Hall2009}, but the notation of~\cite{Moussafir2006}.  Also,
  we define generators~$\sigma_\ip$ as clockwise interchanges rather
  than counterclockwise.}  first define for a quantity~$\fc$ the
operators
\begin{equation}
  \pos\fc \ldef \max(\fc,0),\qquad
  \neg\fc \ldef \min(\fc,0).
\end{equation}
After we define
\begin{equation}
  \cc_{\ip-1} = \ac_{\ip-1} - \ac_\ip - \pos{\bc_\ip} + \neg{\bc_{\ip-1}}\,,
  \label{eq:ccdef}
\end{equation}
we can express the update rules for the~$\sigma_\ip$ acting
on~$\abv=(\ac_1,\ldots,\ac_{\nn-2},\bc_1,\ldots,\bc_{\nn-2})$ as
\begin{subequations}
\begin{align}
  \acnew_{\ip-1} &= \ac_{\ip-1} - \pos{\bc_{\ip-1}}
    - \pos{\l(\pos{\bc_\ip} + \cc_{\ip-1}\r)}\,,\\
  \bcnew_{\ip-1} &= \bc_\ip + \neg{\cc_{\ip-1}}\,,\\
  \acnew_\ip &= \ac_\ip - \neg{\bc_\ip}
    - \neg{\l(\neg{\bc_{\ip-1}} - \cc_{\ip-1}\r)}\,,\\
  \bcnew_\ip &= \bc_{\ip-1} - \neg{\cc_{\ip-1}}\,,
\end{align}
\label{eq:ur}%
\end{subequations}
for~$\ip=2,\ldots,\nn-2$.  For this and the following update rules,
all the other unlisted components of~$\abv$ are unchanged under the
action of~$\sigma_\ip$ or~$\sigma_\ip^{-1}$.  The leftmost ($\ip=1$)
and rightmost ($\ip=\nn-1$) generators require special treatment,
having update rules
\begin{subequations}
\begin{align}
  \acnew_1 &= -\bc_1 + \pos{\l(\ac_1 + \pos{\bc_1}\r)}, \\
  \bcnew_1 &= \ac_1 + \pos{\bc_1}\,.
\end{align}
  \label{eq:ur1}
\end{subequations}
for~$\sigma_1$, and
\begin{subequations}
\begin{align}
  \acnew_{\nn-2} &= -\bc_{\nn-2}
  + \neg{\l(\ac_{\nn-2} + \neg{\bc_{\nn-2}}\r)}\,, \\
  \bcnew_{\nn-2} &= \ac_{\nn-2} + \neg{\bc_{\nn-2}}\,.
\end{align}
  \label{eq:urn}
\end{subequations}
for~$\sigma_{\nn-1}$.

We need to give separate update rules for the
generators~$\sigma_\ip^{-1}$.  With the definition
\begin{equation}
  \dc_{\ip-1} = \ac_{\ip-1} - \ac_\ip + \pos{\bc_\ip} - \neg{\bc_{\ip-1}}\,,
  \label{eq:dcdef}
\end{equation}
the update rules for the~$\sigma_\ip^{-1}$ are
\begin{subequations}
\begin{align}
  \acnew_{\ip-1} &= \ac_{\ip-1} + \pos{\bc_{\ip-1}}
    + \pos{\l(\pos{\bc_\ip} - \dc_{\ip-1}\r)},\\
  \bcnew_{\ip-1} &= \bc_\ip - \pos{\dc_{\ip-1}}\,,\\
  \acnew_\ip &= \ac_\ip + \neg{\bc_\ip}
    + \neg{\l(\neg{\bc_{\ip-1}} + \dc_{\ip-1}\r)},\\
  \bcnew_\ip &= \bc_{\ip-1} + \pos{\dc_{\ip-1}}\,,
\end{align}
\label{eq:iur}%
\end{subequations}
for~$\ip=2,\ldots,\nn-2$.  We also have
\begin{subequations}
\begin{align}
  \acnew_1 &= \bc_1 - \pos{\l(\pos{\bc_1}-\ac_1\r)}, \\
  \bcnew_1 &= \pos{\bc_1} - \ac_1\,.
\end{align}
  \label{eq:iur1}
\end{subequations}
for~$\sigma_1^{-1}$, and
\begin{subequations}
\begin{align}
  \acnew_{\nn-2} &= \bc_{\nn-2} - \neg{\l(
  \neg{\bc_{\nn-2}} - \ac_{\nn-2}\r)}\,, \\
  \bcnew_{\nn-2} &=  \neg{\bc_{\nn-2}} - \ac_{\nn-2}\,.
\end{align}
  \label{eq:iurn}
\end{subequations}
for~$\sigma_{\nn-1}^{-1}$.

Update rules of this form are known as \emph{piecewise-linear}: once
the mins and maxes are resolved, what is left is simply a linear
operation.  However, the mins and maxes are what keeps this from being
a simple linear algebra problem and make the braid dynamics so rich.

Here then is a recipe for computing~$\htopb{\nn}$, the topological entropy
of a braid of $\nn$ particle trajectories:

\bigskip

\noindent
\framebox{\begin{minipage}{.98\textwidth}%
\begin{procedure}[Entropy of periodic braid]
\label{proc:1}
\mbox{}\\[-15pt]
\begin{enumerate}
  \item\label{item:1} Start with an arbitrary initial loop, encoded as a
    vector~$\abv$ (Eq.~\eqref{eq:abvdef}); Set~$\per$ to~$0$;
  \item\label{item:2} For each generator in the braid, use the
    appropriate update rule~\eqref{eq:ccdef}--\eqref{eq:iurn} to
    modify~$\abv$;
  \item\label{item:3} Compute the intersection number~$\Nint(\abv)$ using
    Eq.~\eqref{eq:Nint};
  \item\label{item:4} Repeat steps~\ref{item:2}--\ref{item:3} for all
    generators in the braid;
  \item\label{item:5} Add~$1$ to~$\per$;
    Calculate~$\htopb{\nn}=\per^{-1}\log\Nint(\abv)$;
  \item Repeat steps~\ref{item:2} to~\ref{item:5} until~$\htopb{\nn}$
    converges in step~\ref{item:5}.
\end{enumerate}
\end{procedure}
\end{minipage}}

\bigskip

\noindent\textbf{Remarks:}
\begin{itemize}
\item The procedure above assumes that the braid is \emph{periodic},
  \ie, is obtained from periodic orbits of the flow.  In
  Section~\ref{sec:random} we will discuss how the method differs for
  random braids obtained from sampling arbitrary trajectories (which
  don't necessarily repeat).
\item The dimension of~$\htopb{\nn}$ is inverse time, where the unit of time
  is the period over which the braid is repeated.
\item Even though the discussion so far has described~$\abv$ as a
  vector of integers, the initial condition for~$\abv$ in
  step~\ref{item:1} can in practice be chosen to be a random set of
  real numbers.  (This called a~\emph{projectivized} version of the
  coordinates~\cite{Hall2009}.)
\item As is typical of such exponential growth calculation, it is
  possible that the components of~$\abv$ becomes so large that they
  overflow double-precision arithmetic.  In that case standards
  `renormalization' techniques can be used: divide~$\abv$ by a large
  constant~$\Nint_{\text{overflow}}$, but keep track of how many times
  this division was done.  Then add that multiple
  of~$\log\Nint_{\text{overflow}}$ to the logarithm in
  step~\ref{item:5}.  Another option is to use real or integer
  arbitrary precision arithmetic, but this slows down the calculation.
\item We stress that Moussafir's technique for the computation of a
  braid's entropy is extremely rapid compared to previous methods,
  which typically use \emph{train tracks} and the Bestvina--Handel
  algorithm~\cite{Bestvina1995}, or combinatorial
  methods~\cite{Lefranc2006}.  The rapidity arises from the fact that
  the algorithm keeps a bare minimum of information (the vector~$\abv$)
  to express the topology of an arbitrarily long curve.  The
  Bestvina--Handel algorithm, however, gives more information about
  the braid (such as the existence of invariant curves -- see
  Section~\ref{sec:discussion}).
\end{itemize}

\begin{figure}
\begin{center}
\subfigure[]{
  \includegraphics[width=.47\textwidth]{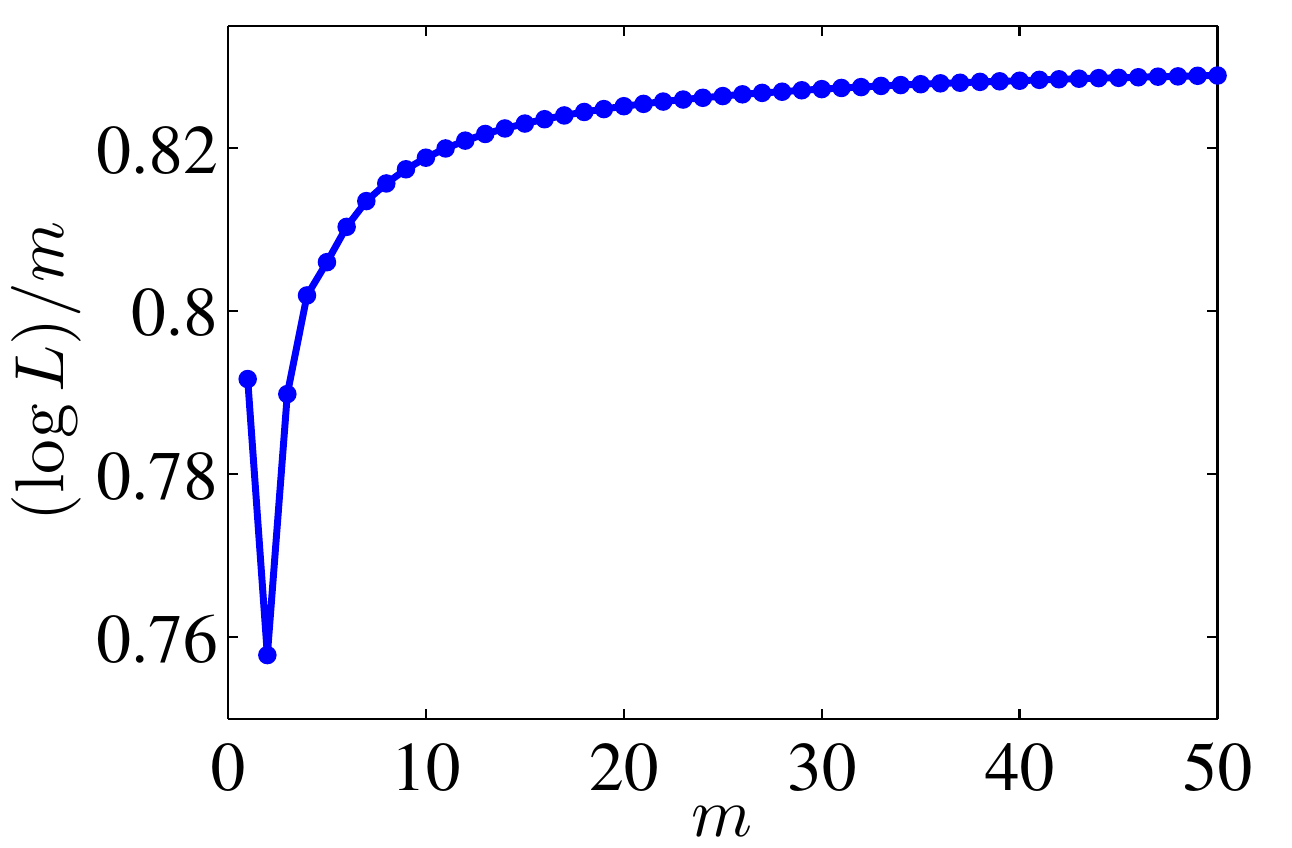}
  \label{fig:dynn_example_L}
}\hspace{1em}
\subfigure[]{
  \includegraphics[width=.45\textwidth]{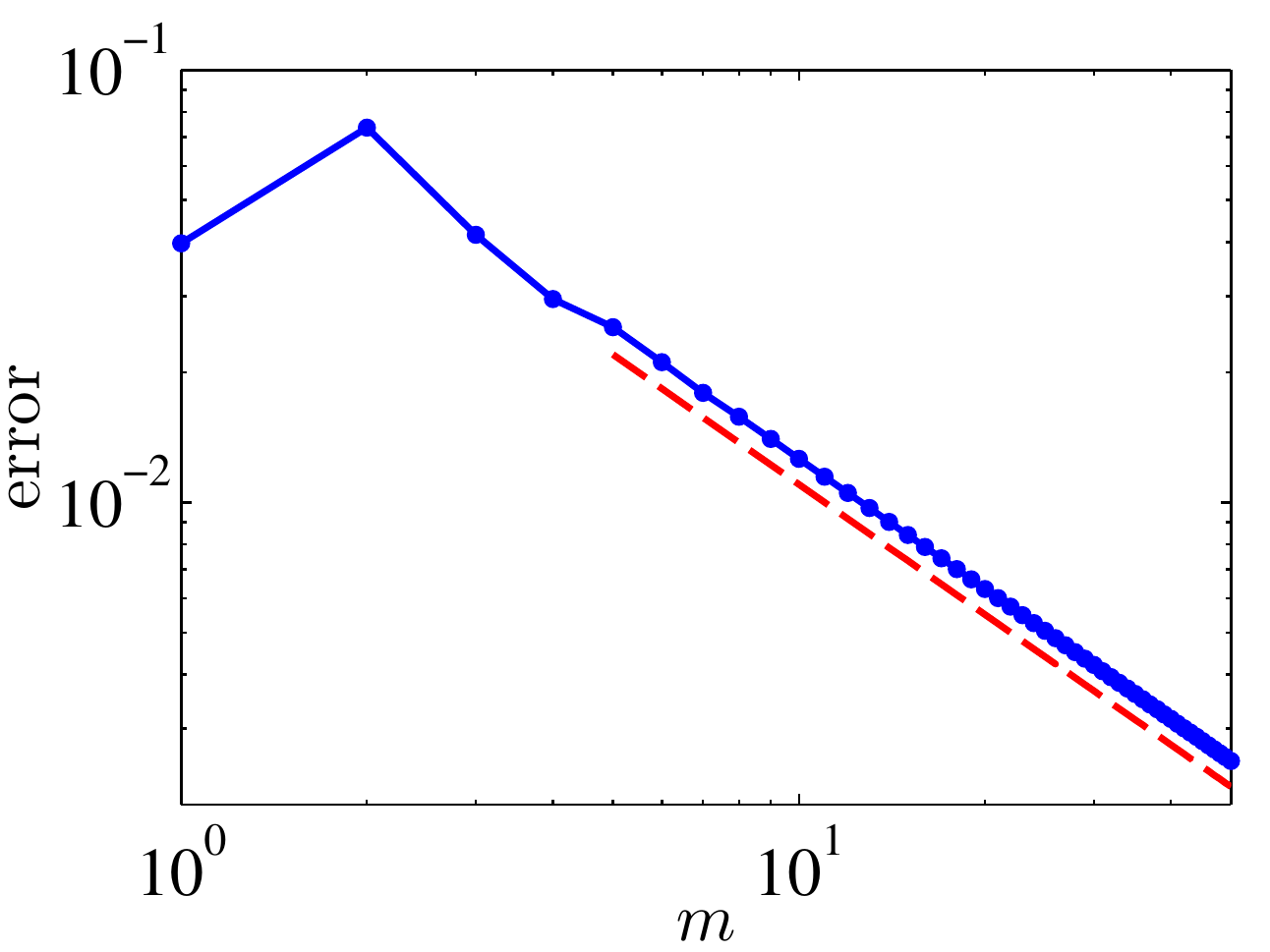}
  \label{fig:dynn_example_error}
}
\end{center}
\caption{For the
  braid~$\sigma_3^{-1}\sigma_2^{-1}\sigma_3^{-1}\sigma_2\sigma_1$
  in Fig.~\ref{fig:braidexample}, (a) Entropy
  estimate~$\per^{-1}\log\Nint(\abv)$ as a function of period~$\per$
  using Procedure~\ref{proc:1}. (b) Error (deviation from the true
  entropy~$\simeq 0.83144$), showing the~$1/\per$ convergence (dashed
  line).}
\end{figure}

The speed of convergence of this procedure is discussed
in~\cite{Moussafir2006,MattFinn2007}.  As an example,
Fig.~\ref{fig:dynn_example_L} shows the result of applying the
procedure to the braid in Fig.~\ref{fig:braidexample}, and
Fig.~\ref{fig:dynn_example_error} shows the convergence rate to the
exact entropy.

\section{Random braids}
\label{sec:random}

From the point of view of data analysis, looking at periodic braid is
not general enough.  Most periodic orbits in a dynamical system are
unstable, and thus they cannot be detected directly.  The trajectories
we have access to are typically chaotic.  Nevertheless, we can ask
what the braid corresponding to a set of orbits tells us about a
dynamical system.  The answer is that its entropy approximates the
`true' topological entropy of the flow, and the approximation gets
better as more particles are added.

There are two ways to analyze random braids generated by chaotic
trajectories: without and with ensemble averaging.  `Without
averaging' means that we have a single realization to study, say~$\nn$
trajectories integrated or measured up to some final time.  Unless the
final time is extremely long, this is not very accurate.  `With
averaging' means that we have the luxury of repeating the experiment
several times, following the same number of trajectories at each
realization (assuming the flow is the same for each realization, at
least in a statistical sense).  We then average over the total number
of realizations of the experiment, in the manner described below.

\subsection{Entropy without averaging}

Let us first describe the procedure without averaging: we assume that
we have obtained a sequence of generators from the trajectories
of~$\nn$ particles, as well as the time at which each crossing occurs
(see Section~\ref{sec:braidsfromflow}).  In the examples presented
here, those trajectories were either computed from randomly-selected
initial conditions, or they were obtained from measured data (the
oceanic floats).

\bigskip

\noindent
\framebox{\begin{minipage}{.98\textwidth}%
\begin{procedure}[Entropy of random braid, without averaging]
\label{proc:2}
\mbox{}\\[-15pt]
\begin{enumerate}
  \item\label{item:r1} Start with an arbitrary initial loop, encoded as a
    vector~$\abv$ (Eq.~\eqref{eq:abvdef});
  \item\label{item:r2} For each generator in the braid, use the
    appropriate update rule~\eqref{eq:ccdef}--\eqref{eq:iurn} to
    modify~$\abv$;
  \item\label{item:r3} Compute the intersection number~$\Nint(\abv)$ using
    Eq.~\eqref{eq:Nint};
  \item\label{item:r5} Plot~$\log\Nint(\abv)$ versus~$\tim$,
    where~$\tim$ is a vector of times when each crossing occurs;
  \item Repeat steps~\ref{item:r2} to~\ref{item:r5} until we can fit a
    line in step~\ref{item:r5}.
\end{enumerate}
\end{procedure}
\end{minipage}
}

\bigskip

\begin{figure}
\begin{center}
\subfigure[]{
  \includegraphics[width=.44\textwidth]{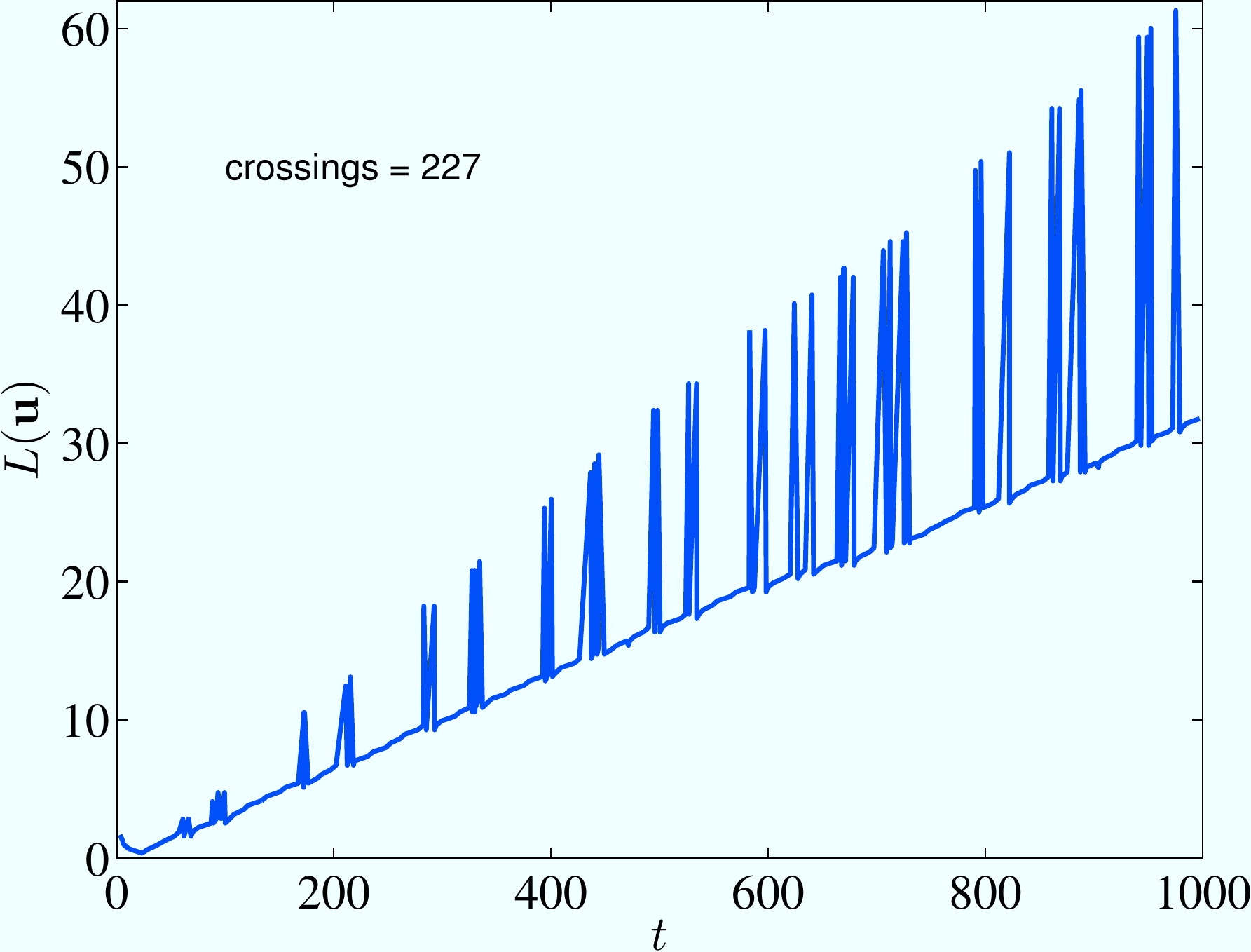}
  \label{fig:blinking_random_regular}
}\hspace{1em}
\subfigure[]{
  \includegraphics[width=.47\textwidth]{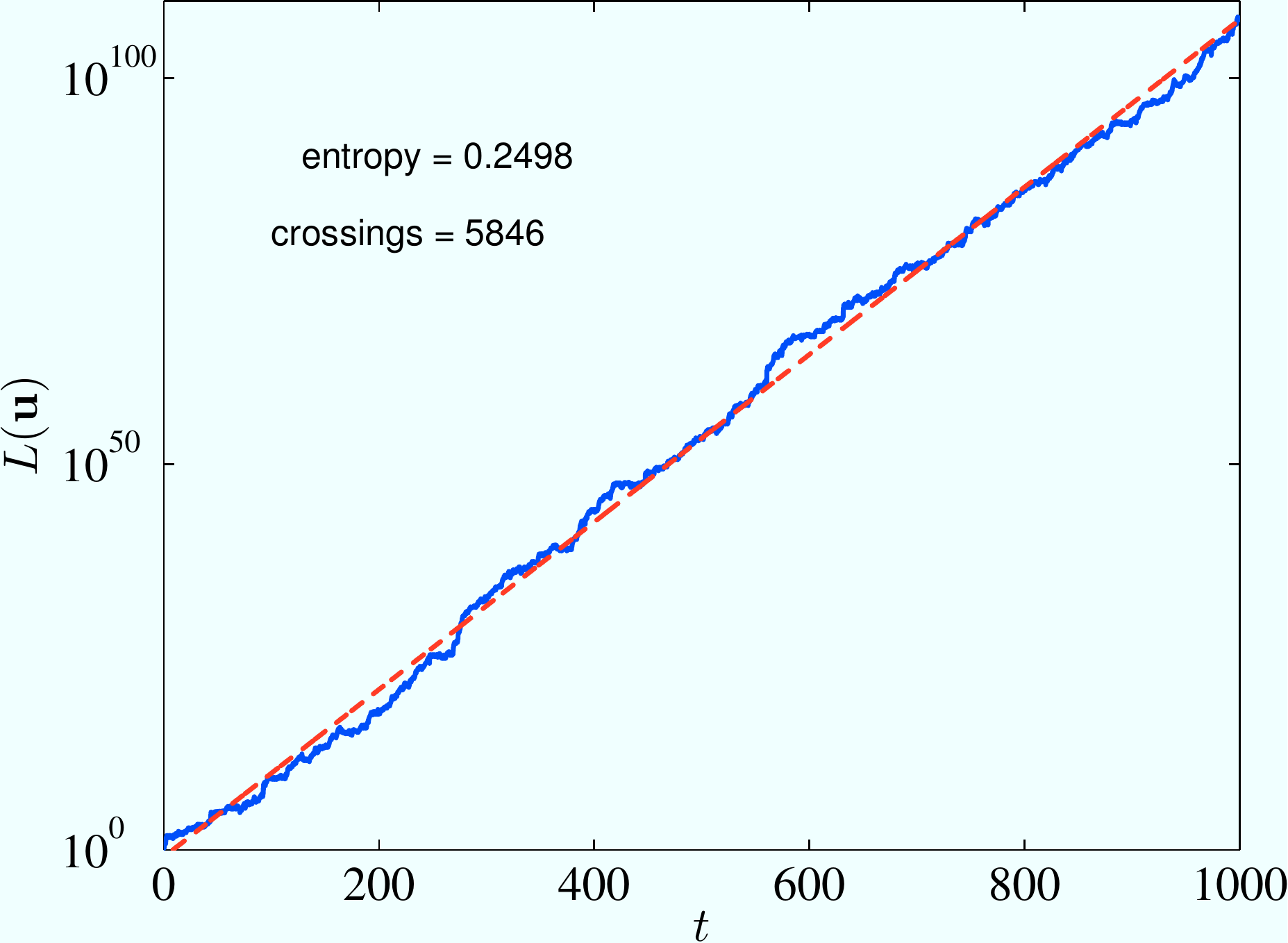}
  \label{fig:blinking_random}
}
\end{center}
\caption{(a) The number of crossings of a loop, versus time, for the
  blinking vortex flow in the regular regime (circulation
  $\Gamma=0.5$, co-rotating vortices; see~\cite{Thiffeault2005}).  The
  growth of~$\Nint(\abv)$ is linear.  (b) Same as in (a), but in the
  chaotic regime (circulation $\Gamma=16.5$, counter-rotating
  vortices; see~\cite{Thiffeault2005}).  The vertical axis is now on a
  log scale; the slope of the line gives the braid entropy
  (Procedure~\ref{proc:2}).}
  \label{fig:blinking_proc2}
\end{figure}

\noindent\textbf{Remarks:}
\begin{itemize}
\item Since the braid is random, we must keep track of the time when
  crossings occur.
\item Most of the remarks from the periodic Procedure~\ref{proc:1} of
  Section~\ref{sec:entrbraid} still apply.
\end{itemize}
Figure~\ref{fig:blinking_proc2} shows an example of applying
Procedure~\ref{proc:2} to~$\nn=3$ particles advected by a blinking
vortex flow~\cite{Aref1984,Thiffeault2005,Kin2005} in the regular
regime (Fig~\ref{fig:blinking_random_regular}) and chaotic regime
(Fig.~\ref{fig:blinking_random}).  In the first case, the growth
of~$\Nint(\abv)$ is roughly linear, so the entropy is zero.  In the
second case the growth is exponential.  Note that the integration time
is quite long, and in Fig.~\ref{fig:blinking_random}~$\Nint(\abv)$
becomes enormous.  For such long integration time, the fit for the
entropy is good.

\subsection{Entropy with averaging}

To get a more accurate measurement of~$\htopb{\nn}$ for random braids,
ensemble averaging is desirable, if we have that luxury.  To implement
this, we integrate a set of~$\nn$ trajectories~$\Nreal$ times,
randomizing the initial condition for each realization.  We obtain a
list of~$\nn\times\Nreal$ trajectories for times~$0\le\tim\le\tmax$,
from which we compute~$\Nreal$ braids and vectors of crossing times.
To do the averaging, we need to be able to compare~$\Nint(\abv)$ at
the same times for each braid, but since crossings occur at different
times we cannot do this directly.  We instead break up the total time
interval into equal subintervals of length~$\dt$, and for each
subinterval and each realization we record~$\Nint(\abv)$ up to
time~$\ti\dt$, where~$\ti$ is an integer with~$0\le\ti\dt\le\tmax$.
We finally obtain~$\Nreal$ lists of~$[\tmax/\dt]$ (square brackets
denote the integer part) intersection numbers~$\Nint(\abv)$, all
sampled at the same times corresponding to each subinterval.  The
whole procedure is summarized as:

\bigskip

\noindent
\framebox{\begin{minipage}{.98\textwidth}%
\begin{procedure}[Entropy of random braid, with averaging]
\label{proc:3}
\mbox{}\\[-15pt]
\begin{enumerate}
\item Start with~$\Nreal$ lists of intersection numbers~$\Nint(\abv)$
  and their crossing times, generated following
  Procedure~\ref{proc:2}, steps~\ref{item:1}--\ref{item:3};
\item For each realization, record the intersection numbers up to
  fixed times~$\ti\dt$, \hbox{$0\le\ti\dt\le\tmax$};
\item\label{item:avg} At each time~$\ti\dt$, compute the
  average~$\langle\log\Nint(\abv)\rangle$ over all~$\Nreal$
  realizations;
\item Plot~$\langle\log\Nint(\abv)\rangle$ versus $\ti\dt$,
  \hbox{$0\le\ti\dt\le\tmax$}, and fit a line to get~$\htopb{\nn}$.
\end{enumerate}
\end{procedure}
\end{minipage}
}

\bigskip

\begin{figure}
\begin{center}
  \includegraphics[width=.6\textwidth]{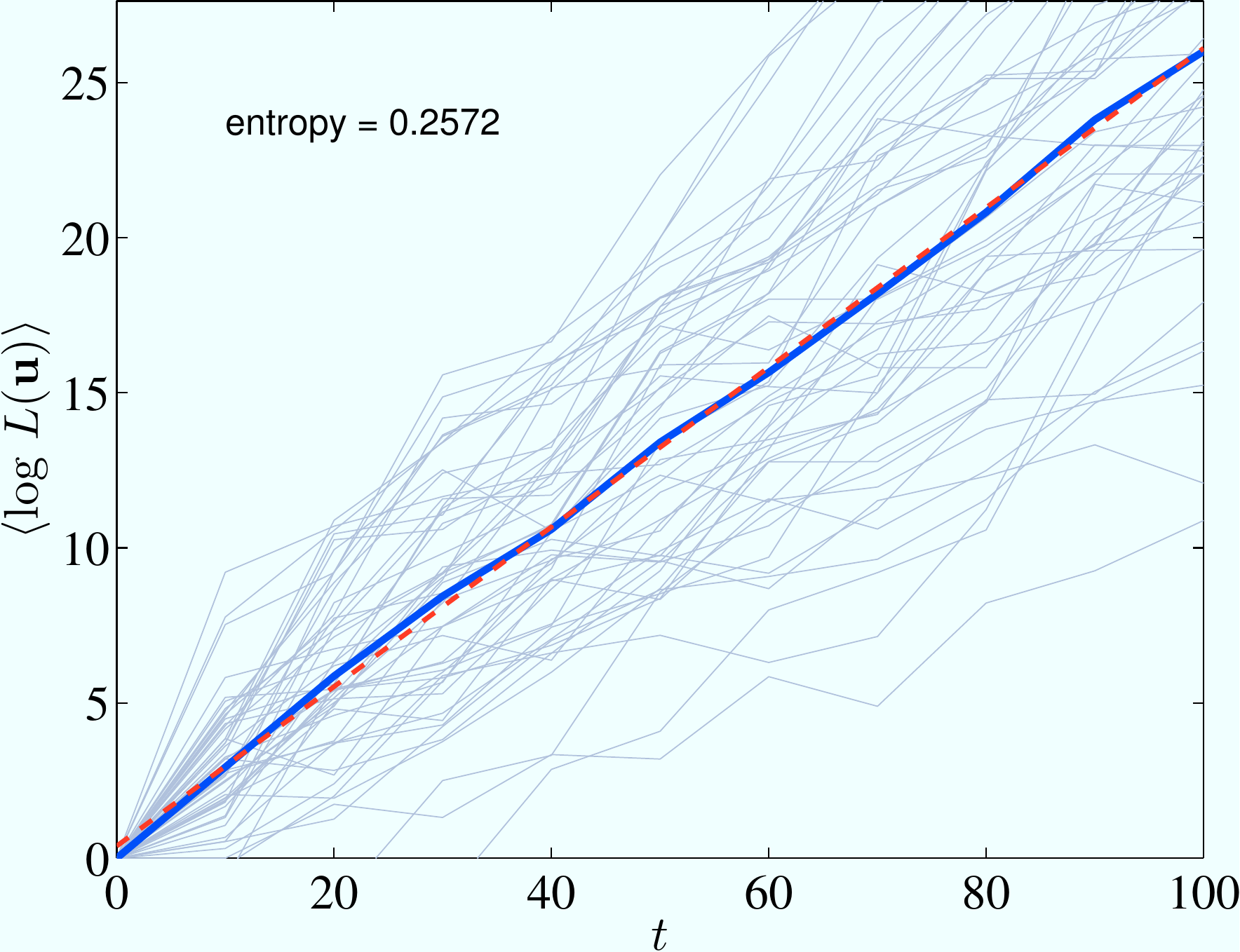}
  \label{fig:blinking_random_average}
\end{center}
\caption{Similar plot to Fig.~\ref{fig:blinking_random}, but after
  averaging~$\langle\log\Nint(\abv)\rangle$ over~$\Nreal=50$ shorter
  trajectories ($\tmax=100$ time units rather than 1000).  The average
  is plotted at each~$\dt=10$ time units (see Procedure~\ref{proc:3}).
  The individual trajectories are shown in the background.}
\end{figure}

\noindent\textbf{Remarks:}
\begin{itemize}
\item We average~$\log\Nint(\abv)$ rather than~$\Nint(\abv)$: not only
  does this ensure that Procedures~\ref{proc:2} and~\ref{proc:3} give
  the same entropy, but it also leads to smaller fluctuations.
\item For best results, the number of subintervals~$[\tmax/\dt]$ has
  to be large enough to get a good fit, but small enough that there
  are several crossings within each subinterval of length~$\dt$.
\end{itemize}
Figure~\ref{fig:blinking_random_average} shows an example of applying
Procedure~\ref{proc:3} with~$\Nreal=50$ realization of~$\nn=3$
particles advected by the same blinking vortex flow as for
Fig.~\ref{fig:blinking_random}.  Notice that the fit is much better,
even though the integration time is shorter.  We used~$[\tmax/dt]=10$
time subintervals of length~$\dt=10$.  An explicit example in Matlab
(for the Duffing oscillator) is given in
Appendix~\ref{code:proc3_example}.

\subsection{Oceanic floats}

As a more practical application of random braids, we consider data for
oceanic floats in the Labrador sea (North Atlantic)~\cite{WFDAC},
discussed in the introduction (Section~\ref{sec:intro_floats}).  The
position of ten floats for a few months is shown in
Fig.~\ref{fig:floats}.  Note that the float trajectories seem more
entangled while they are confined to the Labrador sea (between
Greenland and Labrador), and some eventually escape.  To compute the
braid, we linearly interpolate the float positions to determine when
crossings occur between the~$\nn=10$ floats.  We then use
Procedure~\ref{proc:2} to compute the entropy, as shown in
Fig.~\ref{fig:floats_braid}, since ensemble averaging is not available
here (we only have data for one experiment).  We see in
Fig.~\ref{fig:floats_braid} that~$\Nint(\abv)$ has a convincing
exponential regime for about 150 days, after which floats tend to
escape the Labrador sea and~$\Nint(\abv)$ reaches a plateau.  The
entropy gives us a timescale for the entanglement of floats in the
Labrador sea, here about~$1/0.02 \simeq 50$ days.  This number is easy
to obtain from the raw data: there is no need for a model of the
velocity field.  However, the trajectories need to be long enough for
a significant number of crossings to occur, and localized enough for
particles to actually braid.

More context will be needed to fully understand what it means to say
that the timescale for entanglement is 50 days.  For instance, the
method could be benchmarked by following tracers in simulations of
flows comparable to the Labrador sea, in which braids can be easily
computed.  The measure is also useful for comparing different regions
of the ocean.

Note that there has been previous work on bounding the topological
entropy of experimental data.  \citet{Amon2004} have
obtained lower bounds on entropy and evidence of chaotic behavior in a
nonstationary optical system.  See also the review by \citet{Gilmore1998}.

\section{Discussion}
\label{sec:discussion}

There are two ways to interpret the data obtained from braids of
particles.  The first is to accumulate data for enough particle
trajectories that a good approximation to the topological
entropy~$\htopf$ is obtained.  The drawback to this is that the
convergence of~$\htopb{\nn}$ to~$\htopf$ as~$\nn$ gets larger appears
to be fairly slow~\cite{MattFinn2007}, though more work is needed to
determine this convergence rate.  In this interpretation, the braid
approach is seen as a practical way of measuring~$\htopf$.  This
interpretation is in the same spirit as a Lyapunov exponent.  Its main
advantage is that a single number is easy to comprehend and compare;
its main drawback is that a single number doesn't capture the
subtleties of a particular system.

The second interpretation is to regard~$\htopb{\nn}$ as the
`$\nn$-particle braiding time,' in a similar fashion
that~$\nn$-particle correlation functions are measured.  Thus, the
behavior of~$\htopb{\nn}$ with~$\nn$ carries real information, as it
tells us the typical time for 3 particle trajectories to become
entangled, then 4 particles, etc.  We might call this the `spectrum of
braid entropies' for a dynamical system.  The drawback of this
approach is that it requires a more careful analysis of the data.

The method presented in this paper is limited to two-dimensional
flows.  Indeed, a four-dimensional braid of three-dimensional
particles trajectories is not very useful, as this and all
higher-dimensional braid groups are trivial (strands can always be
disentangled without crossing~\cite{Birman1975}).  The best
alternative is to lift the trajectories of material lines to sheets in
four dimensions, but this presents some daunting visualization
challenges, and there is little developed theory (but
see~\cite{Kamada}).

Finally, the topological entropy~$\htopb{\nn}$ is only the crudest
piece of information that can be extracted from a braid.  Many other
types of invariants can be computed~\cite{Berger2001}.  However, those
invariants don't necessarily have a clear interpretation in terms of
dynamics.  A promising avenue for obtaining much more precise
information on a dynamical system is to find the isotopy class of the
random braid (see Section~\ref{sec:entropy}).  This would tell us, for
instance, whether some floats merely orbit each other and thus behave
as one `trajectory' from the point of view of braiding and entropy.
This is akin to making a braid out of thick rope: even though each
rope is made up of tiny strands, these contribute to the braid as one
large strand.  This sort of approach could help to identify Lagrangian
coherent structures from particle trajectory data, by looking for
decomposable braids.  However, the tools available to do this, such as
the Bestvina--Handel algorithm~\cite{Bestvina1995}, are still slow and
difficult to use on large braids.  A promising approach was recently
used in~\cite{Hall2009}, but needs to be developed further.

\section*{Acknowledgments}

The author thanks Karen Daniels, Tom Solomon, Matt Finn, Matt
Harrington, and Philip Boyland for their help and comments.  This work
was supported by the Division of Mathematical Sciences of the US
National Science Foundation, under grant DMS-0806821.

%\bibliographystyle{jlt}
%\bibliography{bib/journals_abbrev,bib/articles,WFDAC}

\begin{thebibliography}{40}
\newcommand{\enquote}[1]{`#1'}
\providecommand{\natexlab}[1]{#1}
\providecommand{\url}[1]{\texttt{#1}}
\providecommand{\urlprefix}{URL }
\providecommand{\bibinfo}[2]{#2}
\providecommand{\eprint}[2][]{\url{#2}}

\bibitem[{WFD(2004)}]{WFDAC}
\enquote{\bibinfo{title}{{WOCE} subsurface float data assembly center},}
  (\bibinfo{year}{2004}), \bibinfo{note}{\url{http://wfdac.whoi.edu}}.

\bibitem[{{LaCasce}(2008)}]{LacasceAosta2004}
\bibinfo{author}{J.~H. {LaCasce}}, \enquote{\bibinfo{title}{Lagrangian
  statistics from oceanic and atmospheric observations},} in
  \bibinfo{editor}{J.~B. Weiss} and \bibinfo{editor}{A.~Provenzale}, eds.,
  \emph{\bibinfo{booktitle}{Transport and Mixing in Geophysical Flows}},
  \emph{\bibinfo{series}{Lecture Notes in Physics}}, volume
  \bibinfo{volume}{744}, \bibinfo{pages}{165--218}
  (\bibinfo{publisher}{Springer}, \bibinfo{address}{Berlin},
  \bibinfo{year}{2008}).

\bibitem[{Wolf \emph{et~al.}(1985)Wolf, Swift, Swinney, and Vastano}]{Wolf1985}
\bibinfo{author}{A.~Wolf}, \bibinfo{author}{J.~B. Swift},
  \bibinfo{author}{H.~L. Swinney}, and \bibinfo{author}{J.~A. Vastano},
  \enquote{\bibinfo{title}{Determining {L}yapunov exponents from a time
  series},} \emph{\bibinfo{journal}{Physica D}} \textbf{\bibinfo{volume}{16}},
  \bibinfo{pages}{285--317} (\bibinfo{year}{1985}).

\bibitem[{Aref(1984)}]{Aref1984}
\bibinfo{author}{H.~Aref}, \enquote{\bibinfo{title}{Stirring by chaotic
  advection},} \emph{\bibinfo{journal}{J. Fluid Mech.}}
  \textbf{\bibinfo{volume}{143}}, \bibinfo{pages}{1--21}
  (\bibinfo{year}{1984}).

\bibitem[{Boyland \emph{et~al.}(2000)Boyland, Aref, and Stremler}]{Boyland2000}
\bibinfo{author}{P.~L. Boyland}, \bibinfo{author}{H.~Aref}, and
  \bibinfo{author}{M.~A. Stremler}, \enquote{\bibinfo{title}{Topological fluid
  mechanics of stirring},} \emph{\bibinfo{journal}{J. Fluid Mech.}}
  \textbf{\bibinfo{volume}{403}}, \bibinfo{pages}{277--304}
  (\bibinfo{year}{2000}).

\bibitem[{Boyland \emph{et~al.}(2003)Boyland, Stremler, and Aref}]{Boyland2003}
\bibinfo{author}{P.~L. Boyland}, \bibinfo{author}{M.~A. Stremler}, and
  \bibinfo{author}{H.~Aref}, \enquote{\bibinfo{title}{Topological fluid
  mechanics of point vortex motions},} \emph{\bibinfo{journal}{Physica D}}
  \textbf{\bibinfo{volume}{175}}, \bibinfo{pages}{69--95}
  (\bibinfo{year}{2003}).

\bibitem[{Thiffeault(2005)}]{Thiffeault2005}
\bibinfo{author}{J.-L. Thiffeault}, \enquote{\bibinfo{title}{Measuring
  topological chaos},} \emph{\bibinfo{journal}{Phys. Rev. Lett.}}
  \textbf{\bibinfo{volume}{94}}~(\bibinfo{number}{8}), \bibinfo{pages}{084502}
  (\bibinfo{year}{2005}).

\bibitem[{Kin and Sakajo(2005)}]{Kin2005}
\bibinfo{author}{E.~Kin} and \bibinfo{author}{T.~Sakajo},
  \enquote{\bibinfo{title}{Efficient topological chaos embedded in the blinking
  vortex system},} \emph{\bibinfo{journal}{Chaos}}
  \textbf{\bibinfo{volume}{15}}~(\bibinfo{number}{2}), \bibinfo{pages}{023111}
  (\bibinfo{year}{2005}).

\bibitem[{Gouillart \emph{et~al.}(2006)Gouillart, Finn, and
  Thiffeault}]{Gouillart2006}
\bibinfo{author}{E.~Gouillart}, \bibinfo{author}{M.~D. Finn}, and
  \bibinfo{author}{J.-L. Thiffeault}, \enquote{\bibinfo{title}{Topological
  mixing with ghost rods},} \emph{\bibinfo{journal}{Phys. Rev. E}}
  \textbf{\bibinfo{volume}{73}}, \bibinfo{pages}{036311}
  (\bibinfo{year}{2006}).

\bibitem[{Thiffeault \emph{et~al.}(2008)Thiffeault, Finn, Gouillart, and
  Hall}]{Thiffeault2008b}
\bibinfo{author}{J.-L. Thiffeault}, \bibinfo{author}{M.~D. Finn},
  \bibinfo{author}{E.~Gouillart}, and \bibinfo{author}{T.~Hall},
  \enquote{\bibinfo{title}{Topology of chaotic mixing patterns},}
  \emph{\bibinfo{journal}{Chaos}} \textbf{\bibinfo{volume}{18}},
  \bibinfo{pages}{033123} (\bibinfo{year}{2008}).

\bibitem[{Finn \emph{et~al.}(2003)Finn, Cox, and Byrne}]{MattFinn2003}
\bibinfo{author}{M.~D. Finn}, \bibinfo{author}{S.~M. Cox}, and
  \bibinfo{author}{H.~M. Byrne}, \enquote{\bibinfo{title}{Topological chaos in
  inviscid and viscous mixers},} \emph{\bibinfo{journal}{J. Fluid Mech.}}
  \textbf{\bibinfo{volume}{493}}, \bibinfo{pages}{345--361}
  (\bibinfo{year}{2003}).

\bibitem[{Binder and Cox(2008)}]{Binder2008}
\bibinfo{author}{B.~J. Binder} and \bibinfo{author}{S.~M. Cox},
  \enquote{\bibinfo{title}{A mixer design for the pigtail braid},}
  \emph{\bibinfo{journal}{Fluid Dyn. Res.}} \textbf{\bibinfo{volume}{49}},
  \bibinfo{pages}{34--44} (\bibinfo{year}{2008}).

\bibitem[{Vikhansky(2004)}]{Vikhansky2004}
\bibinfo{author}{A.~Vikhansky}, \enquote{\bibinfo{title}{Simulation of
  topological chaos in laminar flows},} \emph{\bibinfo{journal}{Chaos}}
  \textbf{\bibinfo{volume}{14}}~(\bibinfo{number}{1}), \bibinfo{pages}{14--22}
  (\bibinfo{year}{2004}).

\bibitem[{Finn and Thiffeault(2007)}]{MattFinn2007}
\bibinfo{author}{M.~D. Finn} and \bibinfo{author}{J.-L. Thiffeault},
  \enquote{\bibinfo{title}{Topological entropy of braids on the torus},}
  \emph{\bibinfo{journal}{SIAM J. Appl. Dyn. Sys.}}
  \textbf{\bibinfo{volume}{6}}, \bibinfo{pages}{79--98} (\bibinfo{year}{2007}).

\bibitem[{Finn \emph{et~al.}(2006)Finn, Thiffeault, and
  Gouillart}]{MattFinn2006}
\bibinfo{author}{M.~D. Finn}, \bibinfo{author}{J.-L. Thiffeault}, and
  \bibinfo{author}{E.~Gouillart}, \enquote{\bibinfo{title}{Topological chaos in
  spatially periodic mixers},} \emph{\bibinfo{journal}{Physica D}}
  \textbf{\bibinfo{volume}{221}}~(\bibinfo{number}{1}),
  \bibinfo{pages}{92--100} (\bibinfo{year}{2006}).

\bibitem[{Thiffeault and Finn(2006)}]{Thiffeault2006}
\bibinfo{author}{J.-L. Thiffeault} and \bibinfo{author}{M.~D. Finn},
  \enquote{\bibinfo{title}{Topology, braids, and mixing in fluids},}
  \emph{\bibinfo{journal}{Phil. Trans. R. Soc. Lond. A}}
  \textbf{\bibinfo{volume}{364}}, \bibinfo{pages}{3251--3266}
  (\bibinfo{year}{2006}).

\bibitem[{Stremler and Chen(2007)}]{Stremler2007}
\bibinfo{author}{M.~A. Stremler} and \bibinfo{author}{J.~Chen},
  \enquote{\bibinfo{title}{Generating topological chaos in lid-driven cavity
  flow},} \emph{\bibinfo{journal}{Phys. Fluids}} \textbf{\bibinfo{volume}{19}},
  \bibinfo{pages}{103602} (\bibinfo{year}{2007}).

\bibitem[{Vikhansky(2003)}]{Vikhansky2003}
\bibinfo{author}{A.~Vikhansky}, \enquote{\bibinfo{title}{Chaotic advection of
  finite-size bodies in a cavity flow},} \emph{\bibinfo{journal}{Phys. Fluids}}
  \textbf{\bibinfo{volume}{15}}~(\bibinfo{number}{7}),
  \bibinfo{pages}{1830--1836} (\bibinfo{year}{2003}).

\bibitem[{Berger(1990)}]{Berger1990}
\bibinfo{author}{M.~A. Berger}, \enquote{\bibinfo{title}{Third order invariants
  of randomly braided curves},} in \bibinfo{editor}{H.~K. Moffatt} and
  \bibinfo{editor}{A.~Tsinober}, eds., \emph{\bibinfo{booktitle}{Topological
  Fluid Mechanics}}, \bibinfo{pages}{440--448} (\bibinfo{publisher}{Cambridge
  University Press}, \bibinfo{address}{Cambridge, U.K.}, \bibinfo{year}{1990}).

\bibitem[{Nechaev(1996)}]{Nechaev}
\bibinfo{author}{S.~K. Nechaev}, \emph{\bibinfo{title}{Statistics of Knots and
  Entangled Random Walks}} (\bibinfo{publisher}{World Scientific},
  \bibinfo{address}{Singapore; London}, \bibinfo{year}{1996}).

\bibitem[{Artin(1947)}]{Artin1947}
\bibinfo{author}{E.~Artin}, \enquote{\bibinfo{title}{Theory of braids},}
  \emph{\bibinfo{journal}{Ann. Math.}}
  \textbf{\bibinfo{volume}{48}}~(\bibinfo{number}{1}),
  \bibinfo{pages}{101--126} (\bibinfo{year}{1947}).

\bibitem[{Birman \emph{et~al.}(1998)Birman, Ko, and Lee}]{Birman1998}
\bibinfo{author}{J.~S. Birman}, \bibinfo{author}{K.~H. Ko}, and
  \bibinfo{author}{S.~J. Lee}, \enquote{\bibinfo{title}{A new approach to the
  word and conjugacy problems in the braid groups},}
  \emph{\bibinfo{journal}{Adv. Math.}} \textbf{\bibinfo{volume}{139}},
  \bibinfo{pages}{322--353} (\bibinfo{year}{1998}).

\bibitem[{Birman and Brendle(2005)}]{Birman2004}
\bibinfo{author}{J.~S. Birman} and \bibinfo{author}{T.~E. Brendle},
  \enquote{\bibinfo{title}{Braids: {A} survey},} in
  \bibinfo{editor}{W.~Menasco} and \bibinfo{editor}{M.~Thistlethwaite}, eds.,
  \emph{\bibinfo{booktitle}{Handbook of Knot Theory}}
  (\bibinfo{publisher}{Elsevier}, \bibinfo{address}{Amsterdam},
  \bibinfo{year}{2005}), \bibinfo{note}{available at
  \url{http://arXiv.org/abs/math.GT/0409205}}.

\bibitem[{Dynnikov and Wiest(2007)}]{Dynnikov2007}
\bibinfo{author}{I.~A. Dynnikov} and \bibinfo{author}{B.~Wiest},
  \enquote{\bibinfo{title}{On the complexity of braids},}
  (\bibinfo{year}{2007}),
  \bibinfo{eprint}{\url{http://arXiv.org/abs/math.GT/0403177}}.

\bibitem[{Gambaudo and P\'{e}cou(1999)}]{Gambaudo1999}
\bibinfo{author}{J.-M. Gambaudo} and \bibinfo{author}{E.~E. P\'{e}cou},
  \enquote{\bibinfo{title}{Dynamical cocycles with values in the {A}rtin braid
  group},} \emph{\bibinfo{journal}{Ergod. Th. Dynam. Sys.}}
  \textbf{\bibinfo{volume}{19}}, \bibinfo{pages}{627--641}
  (\bibinfo{year}{1999}).

\bibitem[{Lefranc(2006)}]{Lefranc2006}
\bibinfo{author}{M.~Lefranc}, \enquote{\bibinfo{title}{Alternative determinism
  principle for topological analysis of chaos},} \emph{\bibinfo{journal}{Phys.
  Rev. E}} \textbf{\bibinfo{volume}{74}}, \bibinfo{pages}{035202(R)}
  (\bibinfo{year}{2006}).

\bibitem[{Fathi \emph{et~al.}(1979)Fathi, Laundenbach, and
  Po\'{e}naru}]{Fathi1979}
\bibinfo{author}{A.~Fathi}, \bibinfo{author}{F.~Laundenbach}, and
  \bibinfo{author}{V.~Po\'{e}naru}, \enquote{\bibinfo{title}{Travaux de
  {T}hurston sur les surfaces},} \emph{\bibinfo{journal}{Ast\'{e}risque}}
  \textbf{\bibinfo{volume}{66-67}}, \bibinfo{pages}{1--284}
  (\bibinfo{year}{1979}).

\bibitem[{Thurston(1988)}]{Thurston1988}
\bibinfo{author}{W.~P. Thurston}, \enquote{\bibinfo{title}{On the geometry and
  dynamics of diffeomorphisms of surfaces},} \emph{\bibinfo{journal}{Bull. Am.
  Math. Soc.}} \textbf{\bibinfo{volume}{19}}, \bibinfo{pages}{417--431}
  (\bibinfo{year}{1988}).

\bibitem[{Casson and Bleiler(1988)}]{Casson1988}
\bibinfo{author}{A.~J. Casson} and \bibinfo{author}{S.~A. Bleiler},
  \emph{\bibinfo{title}{Automorphisms of surfaces after {N}ielsen and
  {T}hurston}}, \emph{\bibinfo{series}{London Mathematical Society Student
  Texts}}, volume~\bibinfo{volume}{9} (\bibinfo{publisher}{Cambridge University
  Press}, \bibinfo{address}{Cambridge}, \bibinfo{year}{1988}).

\bibitem[{Boyland(1994)}]{Boyland1994}
\bibinfo{author}{P.~L. Boyland}, \enquote{\bibinfo{title}{Topological methods
  in surface dynamics},} \emph{\bibinfo{journal}{Topology Appl.}}
  \textbf{\bibinfo{volume}{58}}, \bibinfo{pages}{223--298}
  (\bibinfo{year}{1994}).

\bibitem[{Birman(1975)}]{Birman1975}
\bibinfo{author}{J.~S. Birman}, \emph{\bibinfo{title}{Braids, Links, and
  Mapping Class Groups}}, Annals of Mathematics Studies
  (\bibinfo{publisher}{Princeton University Press},
  \bibinfo{address}{Princeton, NJ}, \bibinfo{year}{1975}).

\bibitem[{Newhouse and Pignataro(1993)}]{Newhouse1993}
\bibinfo{author}{S.~E. Newhouse} and \bibinfo{author}{T.~Pignataro},
  \enquote{\bibinfo{title}{On the estimation of topological entropy},}
  \emph{\bibinfo{journal}{J. Stat. Phys.}}
  \textbf{\bibinfo{volume}{72}}~(\bibinfo{number}{5-6}),
  \bibinfo{pages}{1331--1351} (\bibinfo{year}{1993}).

\bibitem[{Moussafir(2006)}]{Moussafir2006}
\bibinfo{author}{J.-O. Moussafir}, \enquote{\bibinfo{title}{On the entropy of
  braids},} \emph{\bibinfo{journal}{Func. Anal. and Other Math.}}
  \textbf{\bibinfo{volume}{1}}~(\bibinfo{number}{1}), \bibinfo{pages}{43--54}
  (\bibinfo{year}{2006}).

\bibitem[{Dynnikov(2002)}]{Dynnikov2002}
\bibinfo{author}{I.~A. Dynnikov}, \enquote{\bibinfo{title}{On a {Yang--Baxter}
  map and the {D}ehornoy ordering},} \emph{\bibinfo{journal}{Russian Math.
  Surveys}} \textbf{\bibinfo{volume}{57}}~(\bibinfo{number}{3}),
  \bibinfo{pages}{592--594} (\bibinfo{year}{2002}).

\bibitem[{Hall and Yurtta\c{s}(2009)}]{Hall2009}
\bibinfo{author}{T.~Hall} and \bibinfo{author}{S.~{\"{O}}. Yurtta\c{s}},
  \enquote{\bibinfo{title}{On the topological entropy of families of braids},}
  \emph{\bibinfo{journal}{Topology Appl.}}
  \textbf{\bibinfo{volume}{156}}~(\bibinfo{number}{8}),
  \bibinfo{pages}{1554--1564} (\bibinfo{year}{2009}).

\bibitem[{Bestvina and Handel(1995)}]{Bestvina1995}
\bibinfo{author}{M.~Bestvina} and \bibinfo{author}{M.~Handel},
  \enquote{\bibinfo{title}{Train-tracks for surface homeomorphisms},}
  \emph{\bibinfo{journal}{Topology}}
  \textbf{\bibinfo{volume}{34}}~(\bibinfo{number}{1}),
  \bibinfo{pages}{109--140} (\bibinfo{year}{1995}).

\bibitem[{Amon and Lefranc(2004)}]{Amon2004}
\bibinfo{author}{A.~Amon} and \bibinfo{author}{M.~Lefranc},
  \enquote{\bibinfo{title}{Topological signature of deterministic chaos in
  short nonstationary signals from an optical parametric oscillator},}
  \emph{\bibinfo{journal}{Phys. Rev. Lett.}}
  \textbf{\bibinfo{volume}{92}}~(\bibinfo{number}{9}), \bibinfo{pages}{094101}
  (\bibinfo{year}{2004}).

\bibitem[{Gilmore(1998)}]{Gilmore1998}
\bibinfo{author}{R.~Gilmore}, \enquote{\bibinfo{title}{Topological analysis of
  chaotic dynamical systems},} \emph{\bibinfo{journal}{Rev. Mod. Phys.}}
  \textbf{\bibinfo{volume}{70}}~(\bibinfo{number}{4}),
  \bibinfo{pages}{1455--1529} (\bibinfo{year}{1998}).

\bibitem[{Kamada(2002)}]{Kamada}
\bibinfo{author}{S.~Kamada}, \emph{\bibinfo{title}{Braid and Knot Theory in
  Dimension Four}}, Mathematical Surveys \& Monographs
  (\bibinfo{publisher}{American Mathematical Society}, \bibinfo{year}{2002}).

\bibitem[{Berger(2001)}]{Berger2001}
\bibinfo{author}{M.~A. Berger}, \enquote{\bibinfo{title}{Topological invariants
  in braid theory},} \emph{\bibinfo{journal}{Lett. Math. Phys.}}
  \textbf{\bibinfo{volume}{55}}~(\bibinfo{number}{3}),
  \bibinfo{pages}{181--192} (\bibinfo{year}{2001}).

%\bibitem[{EPAPS(2009)}]{EPAPS}
%\bibinfo{note}{See EPAPS supplementary
%  material at [URL will be inserted by AIP] for the source code of
%  the Matlab files in this appendix}.


\end{thebibliography}

\appendix

\section{Matlab Example Programs}

\newcommand{\codefontsize}{\scriptsize}

%The source code of the Matlab files in this appendix is available
%online~\cite{EPAPS}.
These Matlab programs can be obtained by downloading the source of
this paper at \url{http://arXiv.org/abs/0906.3647}.

\subsection{\texttt{gencross} and \texttt{interpcross}}
\label{code:gencross}

The function \texttt{gencross} computes the generators and crossing
times for particle trajectories (see
Section~\ref{sec:braidsfromflow}).  The function \texttt{interpcross}
is a helper function to \texttt{gencross} that interpolates
crossings.  Both these functions are simple implementations with few
bells and whistles: \texttt{gencross} deals with two or three
adjacent particles crossing between successive timesteps, but it does
not attempt to refine the trajectory (by interpolation or integration)
to resolve crossings.  If it gets confused because too many crossings
are occurring between two successive timesteps, there is not other
option but to refine the data further.

\makebox[\textwidth]{\hrulefill}
{\codefontsize
\begin{verbatim}
function [varargout] = gencross(t,X,Y)
%GENCROSS   Find braid generators from crossings of trajectories.
%   G = GENCROSS(T,X,Y) finds the braid group generators associated with
%   crossings of particle trajectories.  Here T is a column vector of times,
%   and X and Y are coordinates of particles at those times.  X and Y have
%   the same number of rows as T, and N columns, where N is the number of
%   particles.  A projection on the X axis is used to define crossings.
%
%   [G,TC] = GENCROSS(T,X,Y) also returns a vector of times TC when the
%   crossings occurred.

% Find the permutation at each time.
[Xperm,Iperm] = sort(X,2);
dperm = diff(Iperm,1);     % Crossings occur when the permutation changes.
icr = find(any(dperm,2));  % Index of crossings.
gen = []; tcr = [];

for i = 1:length(icr)
  % Order (from left to right) of particles involved in crossing.
  igen = find(dperm(icr(i),:));
  j = 1;
  while j < length(igen)
    if ~sum(dperm(icr(i),igen(j:j+1)))
      %
      % Crossing involves a pair of particles.
      %
      p = Iperm(icr(i),igen(j:j+1));  % The two particles involved in crossing.
      [tt,dY] = interpcross(t,X,Y,icr(i),p(1),p(2));
      tcr = [tcr;tt]; gen = [gen; igen(j)*dY];
      j = j+2;

    elseif ~sum(dperm(icr(i),igen(j:j+2)))
      %
      % Crossing involves a triplet of particles.
      % Two cases are possible:
      %
      if Iperm(icr(i),igen(j)) == Iperm(icr(i)+1,igen(j)+1)
        % Case 1: ABC -> CAB

        % Particles B&C cross first
        p = Iperm(icr(i),igen([j+1 j+2]));
        [tt,dY] = interpcross(t,X,Y,icr(i),p(1),p(2));
        tcr = [tcr;tt]; gen = [gen; igen(j+1)*dY];

        % Particles A&C cross second
        p = Iperm(icr(i),igen([j j+2]));
        [tt,dY] = interpcross(t,X,Y,icr(i),p(1),p(2));
        tcr = [tcr;tt]; gen = [gen; igen(j)*dY];
      elseif Iperm(icr(i),igen(j)) == Iperm(icr(i)+1,igen(j)+2)
        % Case 2: ABC -> BCA

        % Particles A&B cross first
        p = Iperm(icr(i),igen([j j+1]));
        [tt,dY] = interpcross(t,X,Y,icr(i),p(1),p(2));
        tcr = [tcr;tt]; gen = [gen; igen(j)*dY];

        % Particles A&C cross second
        p = Iperm(icr(i),igen([j j+2]));
        [tt,dY] = interpcross(t,X,Y,icr(i),p(1),p(2));
        tcr = [tcr;tt]; gen = [gen; igen(j+1)*dY];
      else
        error('something''s wrong with triple crossing -- increase resolution')
      end
      j = j+3;
    else
      error('too many simultaneous crossings -- increase resolution')
    end
  end
end
varargout{1} = gen;
if nargout > 1, varargout{2} = tcr; end
\end{verbatim}}

\makebox[\textwidth]{\hrulefill}
{\codefontsize
\begin{verbatim}
function [tc,dY] = interpcross(t,X,Y,itc,p1,p2)
%INTERPCROSS   Interpolate a crossing.
%   [TC,DY] = INTERPCROSS(T,X,Y,ITC,P1,P2) is a helper function for
%   GENCROSS.  The input is the data T,X,Y (described in the help for
%   GENCROSS); the index ITC of the time of crossing (i.e., the particles
%   cross between T(ITC) and T(ITC+1); and the indices P1 and P2 of the two
%   particles that are crossing.  INTERPCROSS returns the interpolated
%   crossing time TC, as well as DY (the sign of the difference in Y
%   coordinates) which determines the sign of the generator.

% Refine crossing time and position (linear interpolation).
dt = t(itc+1) - t(itc);  % Time interval.
% Particle velocities in that interval.
U1 = (X(itc+1,p1) - X(itc,p1)) / dt;
V1 = (Y(itc+1,p1) - Y(itc,p1)) / dt;
U2 = (X(itc+1,p2) - X(itc,p2)) / dt;
V2 = (Y(itc+1,p2) - Y(itc,p2)) / dt;
% Interpolated crossing time and Y coordinates.
dtc = -(X(itc,p2) - X(itc,p1)) / (U2-U1);
tc = t(itc) + dtc;
Y1c = Y(itc,p1) + dtc*V1;
Y2c = Y(itc,p2) + dtc*V2;
dY = sign(Y1c-Y2c);
% The sign of Y1c-Y2c determines if the crossing is g or g^-1.
if dY == 0,
  error('can''t resolve sign of generator -- increase resolution');
end
\end{verbatim}}

\subsection{\texttt{loopsigma} and \texttt{loopinter}}
\label{code:loopsigma}

The function \texttt{loopsigma} applies a sequence of braid group
generators to a loop (Section~\ref{sec:entrbraid},
Eqs.~\eqref{eq:ccdef}--\eqref{eq:iurn}).  The function
\texttt{loopinter} computes~$\Nint(\abv)$, the minimum number of
intersections of a loop with the horizontal axis
(Section~\ref{sec:entrbraid}, Eq.~\eqref{eq:Nint}).

\makebox[\textwidth]{\hrulefill}
{\codefontsize
\begin{verbatim}
function up = loopsigma(ii,u)
%LOOPSIGMA   Act on a loop with a braid group generator sigma.
%   UP = LOOPSIGMA(J,U) acts on the loop U (encoded in Dynnikov coordinates)
%   with the braid generator sigma_J, and returns the new loop UP.  J can be
%   a positive or negative integer (inverse generator), and can be specified
%   as a vector, in which case all the generators are applied to the loop
%   sequentially from left to right.

n = length(u)/2 + 2;
a = u(1:n-2); b = u((n-1):end);
ap = a; bp = b;
pos = @(x)max(x,0); neg = @(x)min(x,0);
for j = 1:length(ii)
  i = abs(ii(j));
  if ii(j) > 0
    switch(i)
     case 1
      bp(1) = a(1) + pos(b(1));
      ap(1) = -b(1) + pos(bp(1));
     case n-1
      bp(n-2) = a(n-2) + neg(b(n-2));
      ap(n-2) = -b(n-2) + neg(bp(n-2));
     otherwise
      c = a(i-1) - a(i) - pos(b(i)) + neg(b(i-1));
      ap(i-1) = a(i-1) - pos(b(i-1)) - pos(pos(b(i)) + c);
      bp(i-1) = b(i) + neg(c);
      ap(i) = a(i) - neg(b(i)) - neg(neg(b(i-1)) - c);
      bp(i) = b(i-1) - neg(c);
    end
  elseif ii(j) < 0
    switch(i)
     case 1
      bp(1) = -a(1) + pos(b(1));
      ap(1) = b(1) - pos(bp(1));
     case n-1
      bp(n-2) = -a(n-2) + neg(b(n-2));
      ap(n-2) = b(n-2) - neg(bp(n-2));
     otherwise
      d = a(i-1) - a(i) + pos(b(i)) - neg(b(i-1));
      ap(i-1) = a(i-1) + pos(b(i-1)) + pos(pos(b(i)) - d);
      bp(i-1) = b(i) - pos(d);
      ap(i) = a(i) + neg(b(i)) + neg(neg(b(i-1)) + d);
      bp(i) = b(i-1) + pos(d);
    end
  end
  a = ap; b = bp;
end
up = [ap bp];
\end{verbatim}}

\makebox[\textwidth]{\hrulefill}
{\codefontsize
\begin{verbatim}
function int = loopinter(u)
%LOOPINTER   The number of intersections of a loop with the real axis.
%   I = LOOPINTER(U) computes the minimum number of intersections of a loop
%   (encoded in Dynnikov coordinates) with the real axis.  (See Moussafir
%   (2006), Proposition 4.4.)  U is either a row-vector, or a matrix of
%   row-vectors, in which case the function acts vectorially on each row.

n = size(u,2)/2 + 2;
a = u(:,1:n-2); b = u(:,(n-1):end);
cumb = [zeros(size(u,1),1) cumsum(b,2)];
% The number of intersections before/after the first and last punctures.
% See Hall & Yurttas (2009).
b0 = -max(abs(a) + max(b,0) + cumb(:,1:end-1),[],2); bn = -b0 - sum(b,2);
int = sum(abs(b),2) + sum(abs(a(:,2:end)-a(:,1:end-1)),2) ...
      + abs(a(:,1)) + abs(a(:,end)) + abs(b0) + abs(bn);
\end{verbatim}}

\subsection{\texttt{proc3\_example}}
\label{code:proc3_example}

\begin{figure}
\begin{center}
  \includegraphics[width=.6\textwidth]{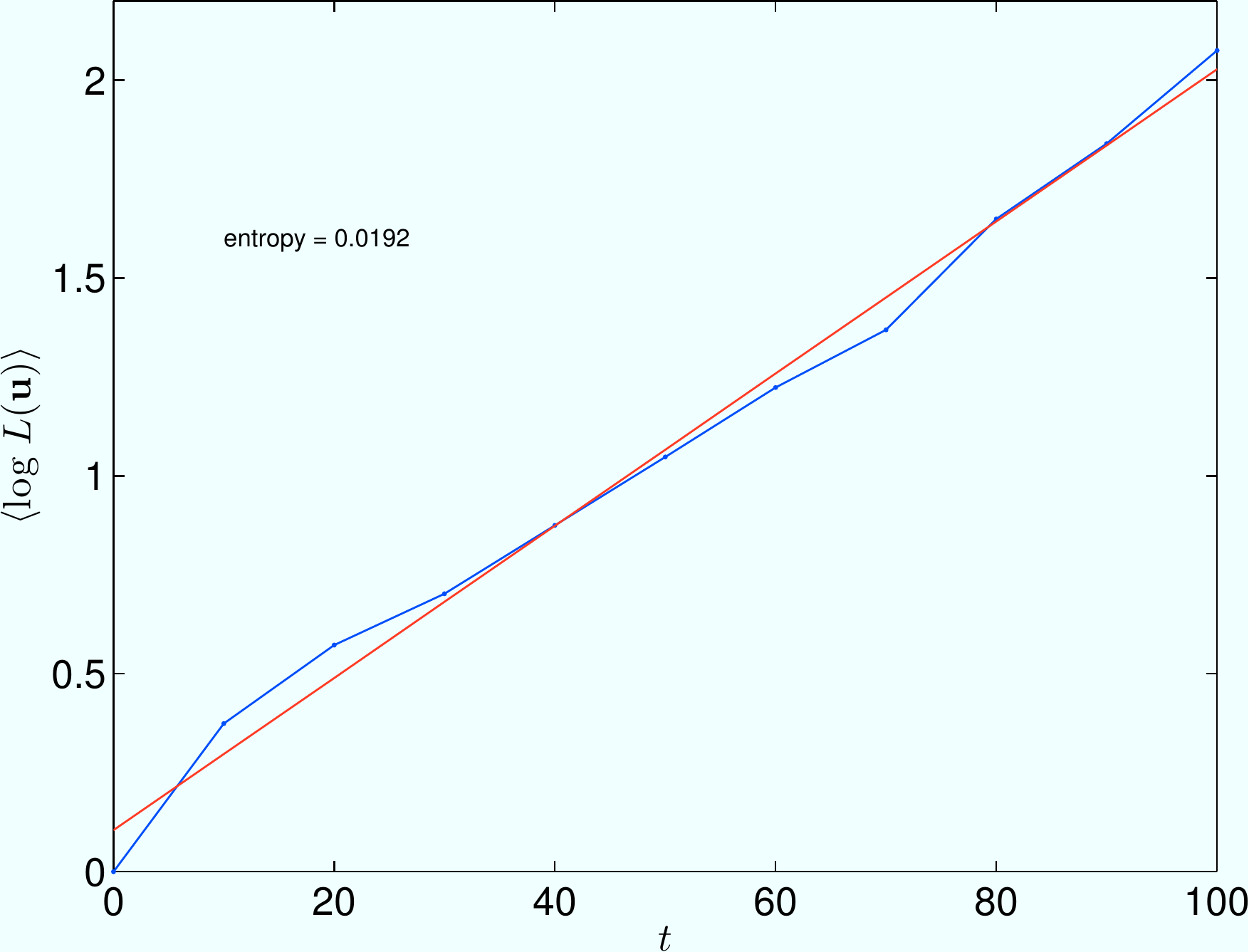}
  \label{fig:proc3_example_fig}
\end{center}
\caption{Output of \texttt{proc3\_example}
  (Section~\ref{code:proc3_example}).}
\end{figure}

The function \texttt{proc3\_example} computes~$\Nint(\abv)$ for a
random braid of 4 particles advected by the Duffing oscillator, using
averaging over 100 realizations (see Section~\ref{sec:random},
Procedure~\ref{proc:3}).

\makebox[\textwidth]{\hrulefill}
{\codefontsize
\begin{verbatim}
function proc3_example

% Procedure 3 example:
% 4 particles advected by the Duffing oscillator, average over realizations.
n = 4; Nreal = 100;
tmax = 100; dt = 10; npts = 3000;
X = zeros(npts,n); Y = zeros(npts,n);
Ll = zeros(Nreal,tmax/dt); tl = dt*(1:tmax/dt);
rand('twister',2);
for r = 1:Nreal
  fprintf(1','Realization %d\n',r)
  % Compute n particle trajectories, from random initial conditions.
  for i = 1:n
    [t,xy] = ode45(@duffing,linspace(0,tmax,npts)',4*rand(1,2)-2);
    X(:,i) = xy(:,1); Y(:,i) = xy(:,2);
  end
  [gen,tc] = gencross(t,X,Y);  % Find generators and crossing times.
  % Act with the generators on a random initial loop.
  up = rand(1,2*(n-2)); up = up/loopinter(up);
  for i = 1:length(gen), up = [up;loopsigma(gen(i),up(end,:))]; end
  % Find the number of intersections with the real axis.
  L{r} = loopinter(up);
  % Keep intersections at a list of fixed time intervals dt, for averaging.
  for q = 1:(tmax/dt)
    idx = find(tc <= tl(q));
    Ll(r,q) = L{r}(idx(end));
  end
end
logLavg = [0 mean(log(Ll),1)]; tl = [0 tl];
m = polyfit(tl,logLavg,1);  % Fit a line
figure(2), plot(tl,logLavg,'b.-'), hold on
text(10,1.6,sprintf('entropy = %.4f',m(1)))
plot(tl,m(1)*tl+m(2),'r-'), hold off
xlabel('t'), ylabel('<log L>')

% -----------------------------------------------------------------
function yt = duffing(t,y)

delta = 1; gamma = 4; omega = 2;
yt = zeros(size(y));
yt(1,:) = y(2,:);
yt(2,:) = y(1,:).*(1 - y(1,:).*y(1,:)) + gamma*cos(omega*t) - delta*y(2,:);
\end{verbatim}}

\end{document}